\documentclass[aps,pra,twocolumn,showpacs,superscriptaddress,preprintnumbers,amsmath,amssymb,footinbib]{revtex4-1}

\usepackage{xcolor}
\usepackage[breaklinks=true,colorlinks,citecolor=blue,linkcolor=blue,urlcolor=blue]{hyperref}

\usepackage{chemformula} 
\usepackage[T1]{fontenc} 
\usepackage{amsmath}
\usepackage{amsfonts}
\usepackage{color}
\usepackage{float}
\usepackage{multirow}
\usepackage{booktabs}
\usepackage{atbegshi}
\usepackage{upgreek}
\DeclareMathSymbol{\shortminus}{\mathbin}{AMSa}{"39}

\usepackage{parskip}
\usepackage{xfp}
\newcommand\SupplementaryMaterials{%
  \xdef\presupfigures{\arabic{figure}}
  \xdef\presuptables{\arabic{table}}
  \xdef\presupsections{\arabic{section}}
  \renewcommand\thefigure{S\fpeval{\arabic{figure}-\presupfigures}}
  \renewcommand\thetable{S\fpeval{\arabic{table}-\presuptables}}
  \renewcommand\thesection{S\fpeval{\arabic{section}-\presupsections}}
}

\begin{document}

\title{Unraveling heat transport and dissipation in suspended \texorpdfstring{MoSe$_2$} ~~crystals from bulk to monolayer}

\author{D.~Saleta~Reig} \affiliation{Catalan Institute of Nanoscience and Nanotechnology (ICN2), BIST and CSIC, Campus UAB, 08193 Bellaterra (Barcelona), Spain}
\author{S.~Varghese} \affiliation{Catalan Institute of Nanoscience and Nanotechnology (ICN2), BIST and CSIC, Campus UAB, 08193 Bellaterra (Barcelona), Spain}
\author{R.~Farris} \affiliation{Catalan Institute of Nanoscience and Nanotechnology (ICN2), BIST and CSIC, Campus UAB, 08193 Bellaterra (Barcelona), Spain}
\author{A.~Block} \affiliation{Catalan Institute of Nanoscience and Nanotechnology (ICN2), BIST and CSIC, Campus UAB, 08193 Bellaterra (Barcelona), Spain}
\author{J.~D.~Mehew} \affiliation{Catalan Institute of Nanoscience and Nanotechnology (ICN2), BIST and CSIC, Campus UAB, 08193 Bellaterra (Barcelona), Spain}
\author{O.~Hellman} \affiliation{Department of Physics, Chemistry and Biology, Linköping University, Sweden}
\author{P.~Wo{\' z}niak}\affiliation{ICFO $-$ Institut de Ciències Fotòniques, Mediterranean Technology Park, Castelldefels (Barcelona) 08860, Spain}
\author{M.~Sledzinska} \affiliation{Catalan Institute of Nanoscience and Nanotechnology (ICN2), BIST and CSIC, Campus UAB, 08193 Bellaterra (Barcelona), Spain}
\author{A.~El Sachat} \affiliation{Catalan Institute of Nanoscience and Nanotechnology (ICN2), BIST and CSIC, Campus UAB, 08193 Bellaterra (Barcelona), Spain}
\author{E.~Chávez-Ángel} \affiliation{Catalan Institute of Nanoscience and Nanotechnology (ICN2), BIST and CSIC, Campus UAB, 08193 Bellaterra (Barcelona), Spain}
\author{S.~O.~Valenzuela} \affiliation{Catalan Institute of Nanoscience and Nanotechnology (ICN2), BIST and CSIC, Campus UAB, 08193 Bellaterra (Barcelona), Spain} \affiliation{ICREA, Pg. Lluís Companys 23, 08010 Barcelona, Spain}
\author{N.~F.~Van Hulst}\affiliation{ICFO $-$ Institut de Ciències Fotòniques, Mediterranean Technology Park, Castelldefels (Barcelona) 08860, Spain} \affiliation{ICREA, Pg. Lluís Companys 23, 08010 Barcelona, Spain}
\author{P.~Ordejón} \affiliation{Catalan Institute of Nanoscience and Nanotechnology (ICN2), BIST and CSIC, Campus UAB, 08193 Bellaterra (Barcelona), Spain}
\author{Z.~Zanolli} \affiliation{Chemistry Department and ETSF, Debye Institute for Nanomaterials Science, Utrecht University, the Netherlands}
\author{C.~M.~Sotomayor~Torres} \affiliation{Catalan Institute of Nanoscience and Nanotechnology (ICN2), BIST and CSIC, Campus UAB, 08193 Bellaterra (Barcelona), Spain} \affiliation{ICREA, Pg. Lluís Companys 23, 08010 Barcelona, Spain}
\author{M.~J.~Verstraete} \affiliation{Nanomat, Q-Mat, CESAM, and European Theoretical Spectroscopy Facility, Université de Liège, B-4000 Liège, Belgium}
\author{K.~J.~Tielrooij} \email[Correspondence to: ]{klaas.tielrooij@icn2.cat} \affiliation{Catalan Institute of Nanoscience and Nanotechnology (ICN2), BIST and CSIC, Campus UAB, 08193 Bellaterra (Barcelona), Spain}

\keywords{heat transport, 2D materials, transition metal dichalcogenides, Raman thermometry, ab initio}

\begin{abstract}
\vspace{0.5cm}
Understanding thermal transport in layered transition metal dichalcogenide (TMD) crystals is crucial for a myriad of applications exploiting these materials. Despite significant efforts, several basic thermal transport properties of TMDs are currently not well understood. Here, we present a combined experimental-theoretical study of the intrinsic lattice thermal conductivity of the representative TMD MoSe$_2$, focusing on the effect of material thickness and the material's environment. We use Raman thermometry measurements on suspended crystals, where we identify and eliminate crucial artefacts, and perform \textit{ab initio} simulations with phonons at finite, rather than zero, temperature. We find that phonon dispersions and lifetimes change strongly with thickness, yet (sub)nanometer thin TMD films exhibit a similar in-plane thermal conductivity ($\sim$20~Wm$^{\shortminus1}$K$^{\shortminus1}$) as bulk crystals ($\sim$40~Wm$^{\shortminus1}$K$^{\shortminus1}$). This is the result of compensating phonon contributions, in particular low-frequency modes with a surprisingly long mean free path of several micrometers that contribute significantly to thermal transport for monolayers. We furthermore demonstrate that out-of-plane heat dissipation to air is remarkably efficient, in particular for the thinnest crystals. These results are crucial for the design of TMD-based applications in thermal management, thermoelectrics and (opto)electronics.
\end{abstract}

\maketitle

\section{Introduction}

Two-dimensional (2D) materials in single or few-layer form have great potential as nanometer thin building blocks for flexible and wearable (opto)electronic and photonic~\cite{wang2012electronics, seo2019fully, manzeli20172d} devices. Concrete examples of promising devices based on 2D transition metal dichalcogenides (TMDs) are photodetectors~\cite{LopezSanchez2013, du2020ultrathin}, transistors~\cite{Roy2014, chuang2016low}, gas sensors~\cite{Yang2017gas, jha2019mose2}, and thermoelectric generators~\cite{oh2016chemically}. Many of these applications rely on the remarkable properties of van der Waals crystals that appear upon reaching, or approaching, the monolayer thickness limit. Examples are the crossover from indirect to direct bandgap at the monolayer limit of MoS$_2$~\cite{mak2010atomically} and other TMDs, a metal-to-semiconductor transition in PtSe$_2$~\cite{ciarrocchi2018thickness}, mechanical softening of MoSe$_2$ films~\cite{babacic2021thickness}, and layer-dependent magnetic phases in CrI$_3$~\cite{Huang2017}. The ability to control the thickness of layered materials allows one to engineer their electrical, optical, mechanical and magnetic properties.

The thermal properties of layered materials have so far received less attention than their electronic and optical counterparts, although several remarkable and exotic thermal transport phenomena have been found. Interesting observations are the ultrahigh in-plane thermal conductivity of graphene~\cite{balandin2008superior} and hexagonal boron nitride (hBN)~\cite{cai2019high}, the highly anisotropic thermal conductivity of TMDs~\cite{jiang2017probing}, and the occurrence of second sound in graphite~\cite{Huberman2019}. However, there are still many open questions concerning the very basic, yet critical, thermal transport properties of TMDs at room temperature~\cite{Zhao2020}. In particular, experimental values of the in-plane lattice thermal conductivity $\kappa$ vary substantially, ranging from 6~Wm$^{\shortminus1}$K$^{\shortminus1}$~\cite{zobeiri2019frequency} to 59~Wm$^{\shortminus1}$K$^{\shortminus1}$~\cite{zhang2015measurement} for MoSe$_2$, and it is not clear how the thermal conductivity changes with the thickness of TMD flakes~\cite{easy2021experimental, bae2017thickness, yuan2018nonmonotonic}. A systematic experimental study with a broad range of thicknesses is lacking. Moreover, the calculated thermal conductivities extracted from atomistic simulations also give scattered results, ranging from 17.6~Wm$^{\shortminus1}$K$^{\shortminus1}$~\cite{C5RA19747C} to 54~Wm$^{\shortminus1}$K$^{\shortminus1}$~\cite{Gu2014dft} for monolayer MoSe$_2$. Also in the theoretical approaches, a systematic thickness variation is lacking, as most studies focused either on monolayer or bulk MoSe$_2$. The effect of the environment on thermal transport in TMDs has furthermore not received much attention, despite that a significant effect was observed for graphene~\cite{chen2011raman}. This situation for MoSe$_2$ is representative for all layered materials in the TMD family~\cite{Zhao2020}.

Performing reliable experimental and theoretical thermal transport studies over a broad thickness range, down to the molecular monolayer, is challenging. Experimental approaches can be susceptible to thickness-dependent artefacts, and require reproducible fabrication of a large number of clean samples with the required thicknesses. Theoretical approaches based on molecular dynamics simulations are limited in accuracy by the choice of empirical interatomic potentials, while \textit{ab initio} simulations often examine phonons at zero temperature, rather than at room temperature, and simulations of thicknesses other than monolayer and bulk are computationally costly.

In this work, we overcome these challenges and reach a deep understanding of thermal transport properties of TMD crystals. In particular, we establish how the in-plane lattice thermal conductivity $\kappa$ depends on crystal thickness, that is, the number of molecular layers. For this, we systematically vary the thickness down to the monolayer limit, both in experiment and simulations. Whereas we focus on MoSe$_2$ crystals, the obtained results are representative for other TMDs. In our experimental approach, we exploit the widely used technique of Raman thermometry, where we identify and eliminate important artefacts that can have a strong influence on the experimentally obtained thermal conductivity. In our theoretical approach, we perform \textit{ab initio} simulations based on density functional theory and Boltzmann transport theory, including anharmonic renormalization yielding accurate results also at room temperature. We employ SIESTA~\cite{Soler_2002, siesta}, which is particularly suitable for atomistic simulations with a large number of atoms, such that we can obtain results up to several molecular layers. 

We find that the main contribution to the in-plane thermal conductivity in few-layer MoSe$_2$ comes from phonon modes centered around 1~THz. Towards the monolayer limit, the contribution of these modes decreases substantially, as there are fewer modes and the phonon lifetimes decrease. These effects are counteracted by the appearance of ``surface'' modes well below 1~THz and with exceptionally long mean free path (MFP) of several micrometers, which contribute substantially to thermal transport. This results in an in-plane thermal conductivity that is mostly constant up to a thickness of $\sim$10 layers, with $\kappa$ $\sim$20~Wm$^{\shortminus1}$K$^{\shortminus1}$ ($\sim$25~Wm$^{\shortminus1}$K$^{\shortminus1}$) according to experiment (simulation), after which it increases progressively towards $\sim$40~Wm$^{\shortminus1}$K$^{\shortminus1}$ for bulk (experiment and simulation). We furthermore find that thermal transport is strongly affected by the material's environment, in particular for monolayer crystals, where $>$80\% of the thermal power is lost through out-of-plane heat dissipation to surrounding molecules. We extract a remarkably large heat transfer coefficient $h_{\rm c}$ up to 30,000 Wm$^{\shortminus2}$K{$^{\shortminus1}$} for monolayer MoSe$_2$.

\section{Results and Discussion}

\subsection*{Experimental approach}

\begin{figure*}[htbp]
\centering
\includegraphics[width=\linewidth]{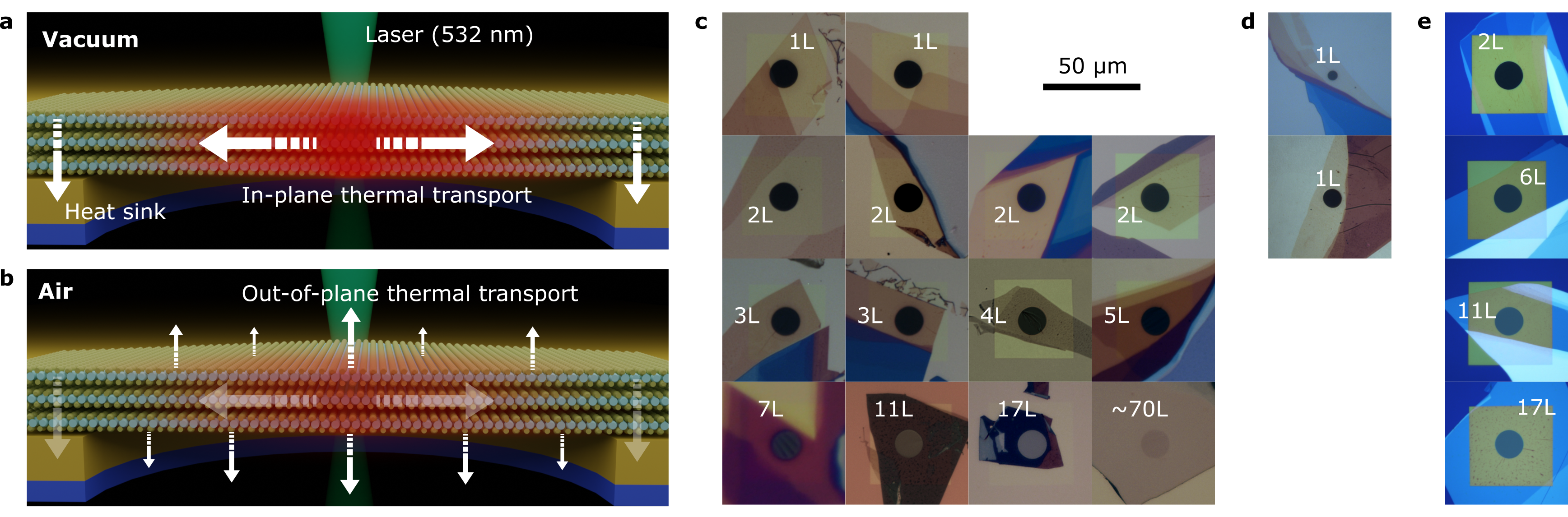}
\caption{\textbf{Concept of the thermal transport experiments and investigated samples.} 
\textbf{a)}~Schematic representation of a suspended trilayer MoSe$_2$ crystal in vacuum, where absorbed 532~nm laser light in the center of the suspended region leads to local heating, and subsequent heat spreading towards the heat sink at the edge of the suspended region, establishing a steady state temperature profile that depends on the in-plane thermal conductivity $\kappa$. 
\textbf{b)}~In air, additional out-of-plane dissipation occurs. 
\textbf{c)}~Optical reflection images of suspended MoSe$_2$ flakes with a thickness varying from monolayer to $\sim$70 layers, suspended over gold-coated substrates with circular holes with a radius of 7.5~$\upmu$m (black/grey central circle), inside Si$_3$N$_4$ membranes (yellow squares). Several flakes have regions with different thicknesses, yet the thickness is uniform in the suspended region in all cases. \textbf{d)}~Images of monolayer flakes that were transferred onto gold-coated substrates with holes with a radius of 2.5 and 5~$\upmu$m, used to study the effect of hole size. \textbf{e)}~Images of flakes that were transferred onto bare 200~nm thick Si$_3$N$_4$ membranes without gold coating, used to study the effect of varying heat sinking properties of the substrate. The 50~$\upmu$m scale bar is common for panels \textbf{(c)}, \textbf{(d)} and~\textbf{(e)}.}
\label{fig:FIG1}
\end{figure*}

One of the most common methods to study thermal properties of thin films is Raman thermometry~\cite{zhang2015measurement, bae2017thickness, yuan2018nonmonotonic, easy2021experimental}, where a laser beam serves both as a heater and a thermometer. The thermometer works via Raman scattering of the laser light, where the frequency shift of a temperature-calibrated Raman mode serves as a probe of the local temperature of a suspended sample. This technique benefits from a relatively simple implementation, contactless nature, and no stringent sample requirements, apart from the presence of a temperature-sensitive Raman active mode. In our experiments (see Methods for details), we use continuous wave (CW) light with a wavelength of 532~nm to heat a local spot with a $1/e$ spot size $r_0$ of $\sim$1~$\upmu$m in the center of a suspended MoSe$_2$ crystal (see Fig.~\ref{fig:FIG1}\textbf{a-b}). Subsequent cooling occurs -- in the ideal situation -- by radial, diffusive flow of heat towards the edge of the suspended region of the crystal, where the substrate acts as a heat sink. We then probe the temperature at the location of the laser spot, corresponding to the steady-state situation where laser-induced heating is compensated by cooling through heat flow and subsequent heat sinking. Thus, a higher (lower) steady-state temperature indicates less (more) efficient cooling, which in turn implies a lower (higher) $\kappa$. For thin exfoliated TMD flakes with high crystallinity the obtained $\kappa$ corresponds to in-plane transport, as the out-of-plane thermal conductivity is typically more than an order of magnitude lower~\cite{jiang2017probing}. 

We use exfoliated MoSe$_2$ crystals suspended over substrates with a circular hole, fabricated using dry transfer, as described in the Methods. This fabrication method leads to crystalline, residue-free, suspended crystals, as shown in Ref.~\cite{Varghese2021b}, which allows us to probe the intrinsic material properties of MoSe$_2$ crystals. We vary the thickness from monolayer (1L) up to $\sim$70 layers (70L), fully covering the 1L to 5L range (see Fig.~\ref{fig:FIG1}\textbf{c$-$e}). This corresponds to a thickness ranging from 0.7~nm up to $\sim$50~nm. We carefully determined these thicknesses using a combination of optical contrast, atomic force microscopy and photoluminescence measurements (see Supporting Fig.~\ref{fig:FIGS1}). Importantly, we use more than one sample with the same thickness in the 1L to 3L regime, including four bilayer samples, in order to assess the reproducibility of both our samples and our experimental technique. We suspend the flakes over circular holes with a radius of 7.5~$\upmu$m, in the centre of 200~nm thick Si$_3$N$_4$ membranes that are coated with a 50~nm thick layer of gold (see Fig.~\ref{fig:FIG1}\textbf{c}). For comparison, we also study monolayer samples suspended over smaller holes (see Fig.~\ref{fig:FIG1}\textbf{d}), and flakes with varying thickness, transferred on Si$_3$N$_4$ substrates without gold coating (see Fig.~\ref{fig:FIG1}\textbf{e}), aimed at understanding the effect of the substrate on the extracted thermal conductivity. 

\begin{figure*}[htbp]
\centering
\includegraphics[width=\linewidth]{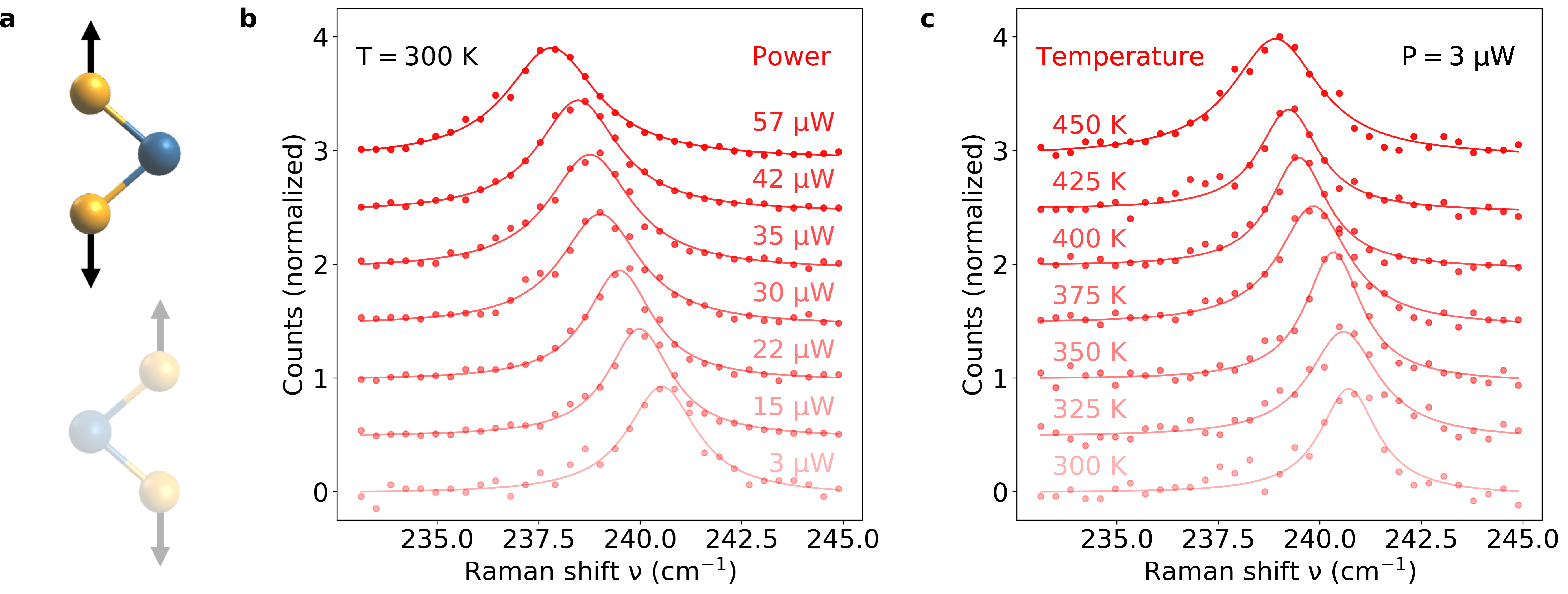}
\caption{\textbf{Raman thermometry of a suspended monolayer MoSe$_2$.} \textbf{a)}~Schematic representation of the A$_{\rm 1g}$ mode of MoSe$_2$. \textbf{b)}~Raman spectra at 532~nm for increasing laser power $P$, showing an increasing red-shift of the A$_{\rm 1g}$ mode due to laser-induced heating. \textbf{c)}~Calibration measurements of Raman spectra at 532~nm for increasing sample temperature of the sample stage, showing an increasing red-shift. Here, the laser power was kept very low, in order to avoid laser-induced heating. For similar measurements on thicker flakes, see Supporting Figs.~\ref{fig:FIG-rawP_vacuum} and~\ref{fig:temp_calib}.}
\label{fig:FIG2}
\end{figure*}

We perform Raman thermometry measurements on all the suspended MoSe$_2$ samples shown in Fig.~\ref{fig:FIG1}\textbf{c$-$e}, exploiting the temperature-sensitive $A_{\rm 1g}$ Raman mode (see Fig.~\ref{fig:FIG2}\textbf{a}). In Fig.~\ref{fig:FIG2}\textbf{b}, we show how the peak frequency of this mode shifts with laser power at the sample position, $P$, for monolayer MoSe$_2$ (see Supporting Fig.~\ref{fig:FIG-rawP_vacuum} for the results for other thicknesses): a higher laser power induces a larger temperature increase $\Delta T$, and therefore a larger red-shift. We correlate the red-shift of the A$_{\rm 1g}$ peak, $\Delta\nu$, with the increase in temperature, $\Delta T$, by measuring the Raman spectrum at very low incident power, while varying the temperature of the crystal using a controlled sample stage (see Fig.~\ref{fig:FIG2}\textbf{c} for monolayer MoSe$_2$, and Supporting Fig.~\ref{fig:temp_calib} for the results for other thicknesses). We find that the temperature coefficients $\chi_{\rm T} = \partial\nu / \partial T$ change from $\shortminus 0.007$~cm$^{\shortminus 1}$/K for bulk to $\shortminus 0.015$~cm$^{\shortminus 1}$/K for monolayer MoSe$_2$ (see Supporting Table~\ref{tab:tab1}). We then use these temperature coefficients to convert the laser-induced red-shift $\Delta \nu$ of the peak of the Raman signal into a local temperature rise that depends on laser power $\Delta T (P)$. 

In order to extract the in-plane thermal conductivity, we perform a linear fit to the extracted $\Delta T$ as a function of absorbed laser power $P_{\rm abs}$, obtaining the slope $\partial T/\partial P_{\rm abs}$, and then use the following equation, following Ref.~\cite{cai2010thermal}:
\begin{equation}
 \kappa = \alpha \cdot \frac{1}{2 \pi d} \cdot \left ( \frac{\partial T}{\partial P_{\rm abs}} \right )^{-1} \cdot \ln\left({\frac{R}{r_0}}\right) \hspace{0.25cm},
\label{eq:1}
\end{equation}
where $R$ is the hole radius, $r_0$ is the laser spot radius, and $\alpha$ is a prefactor that is a function of the ratio $R / r_0$. For our experimental conditions, $\alpha \approx 1$~\cite{cai2010thermal}. This equation for $\kappa$ is valid when the only cooling channel is in-plane diffusive heat transport to the edge of a circular suspended material, where perfect heat-sinking occurs, such that the crystal is at ambient temperature. The accurate extraction of $\kappa$ relies on knowledge of the laser spot size $r_0$ and the optical absorption of each of the flakes, which were measured independently (see Methods). We confirmed the validity of Eq.~(\ref{eq:1}) using a numerical simulation of the Raman thermometry experiment (see Supporting Fig.~\ref{fig:Alex1}).

\begin{figure*}[htbp]
\centering
 \includegraphics[width=\linewidth]{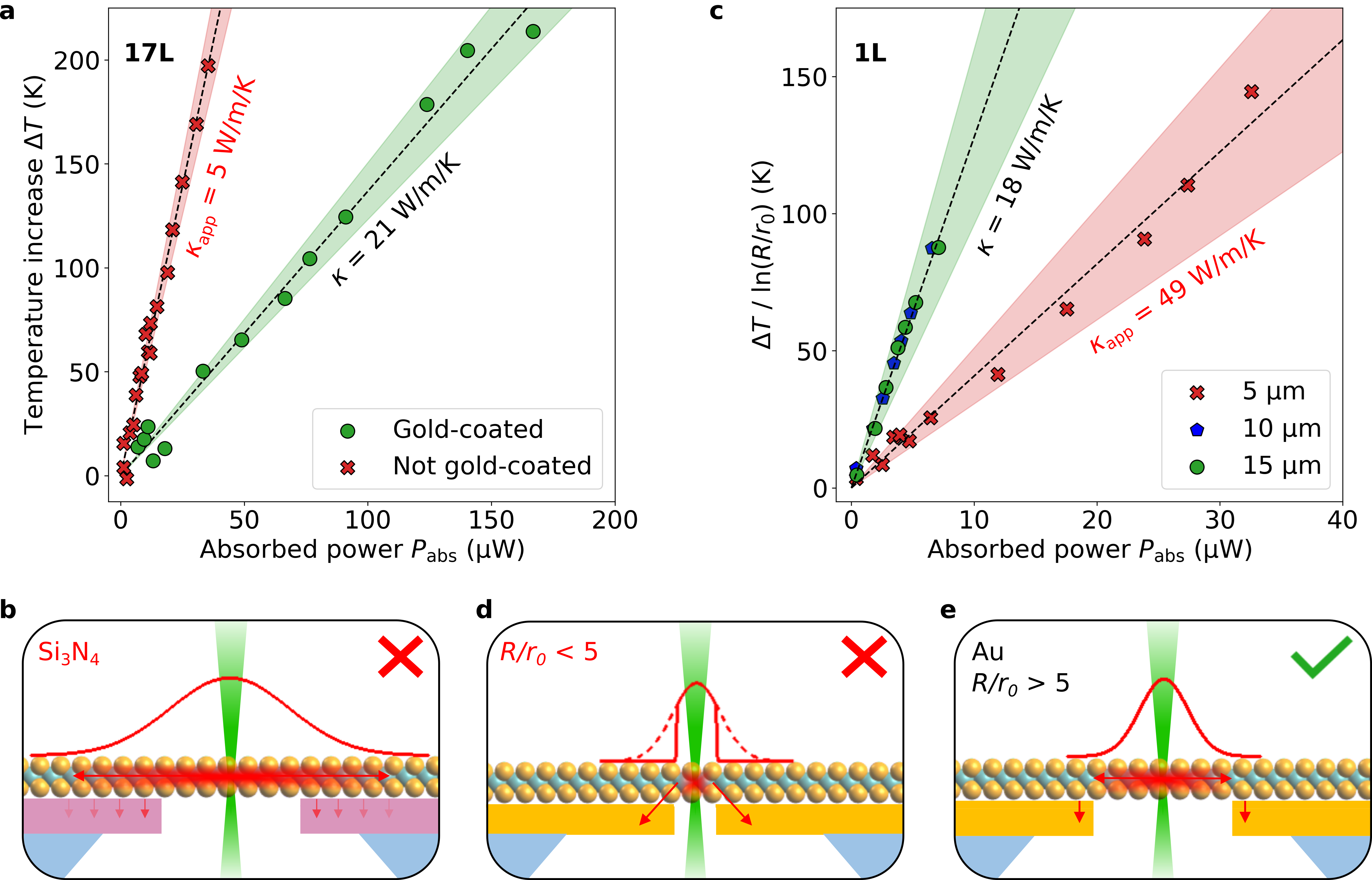}
\caption{\textbf{Artefact identification and elimination.} \textbf{a)}~Temperature increase $\Delta T$ as a function of absorbed power $P_{\rm abs}$ for 17L flakes on substrates without gold coating (red crosses and red shaded area) and with gold coating (green circles and green shaded ares). Gold coating improves heat sinking to the substrate, and therefore gives a more reliable estimate of the intrinsic $\kappa$ of MoSe$_2$.
\textbf{b)}~Side-view schematic of substrate with artefacts due to reduced heat sinking. 
\textbf{c)}~Temperature increase, normalized by $\ln{(R/r_0)}$, as a function of absorbed power $P_{\rm abs}$ for 1L flakes on substrates with hole diameters of 5~$\upmu$m (red crosses), 10~$\upmu$m (blue pentagons) and 15~$\upmu$m (green circles). The smallest hole size leads to an overestimation of $\kappa$, likely due to an overestimation of $P_{\rm abs}$ due to beam clipping on the hole. 
\textbf{d)}~Side-view schematic of substrate with artefacts due to insufficiently large hole size. 
\textbf{e)}~Side-view schematic of artefact-free substrate.
}
\label{fig:FIG3}
\end{figure*}

\subsection*{Experimental approach: eliminating substrate-induced artefacts}

Before presenting the results of the intrinsic thermal conductivity of MoSe$_2$ crystals, we demonstrate how substrate properties can affect the extracted thermal conductivity, leading to a non-intrinsic, apparent, thermal conductivity $\kappa_{\rm app}$ (see Supporting Fig.~\ref{fig:FigS-coating}). In Fig.~\ref{fig:FIG3}\textbf{a}, we compare the results for two suspended flakes with a thickness of 17L placed on Si$_3$N$_4$ substrates with and without 50~nm gold-coating. We observe a much higher $\Delta T$ for the substrate without gold coating, which we ascribe to less efficient heat sinking. We confirm this difference in heat sinking efficiency by measuring, with relatively high incident power, on the substrate-supported regions, where $\Delta T$ is larger for the non-coated substrate than for the gold-coated substrate (see Supporting Fig.~\ref{fig:FigS-coating}\textbf{b}). The main reason for this is likely that the thermal conductivity of gold is larger than that of thin Si$_3$N$_4$ films~\cite{Zhang1995}. Using scanning thermometry, with typical incident laser powers, we furthermore find a significant temperature increase on the suspended region and no visible temperature increase on the supported region (see Supporting Fig.~\ref{fig:FigS-coating}\textbf{c}). We conclude that heat sinking to the gold-coated sample is efficient, implying that the requirements for using Eq.~(\ref{eq:1}) are met, and thus we obtain the intrinsic thermal conductivity $\kappa$. The less efficient heat sinking of the non-coated substrate leads to more heat accumulation on the suspended region of the flake and therefore a reduced apparent thermal conductivity $\kappa_{\rm app}$ when using Eq.~(\ref{eq:1}), as illustrated in Fig.~\ref{fig:FIG3}\textbf{b}. Since in this case the boundary conditions used to arrive at Eq.~(\ref{eq:1}) are not fulfilled, one should use a modified version of Eq.~(\ref{eq:1}), taking into account, for example, the thermal boundary conductance between MoSe$_2$ and the substrate material, and the substrate's thermal conductivity. However, this requires accurate knowledge of such material parameters, which likely introduces additional uncertainty and, possibly, errors in the obtained conductivity. We conclude that it is crucial to use gold-coated substrates, as this leads to efficient heat sinking, such that Eq.~(\ref{eq:1}) is valid. 

To study possible substrate-induced artefacts related to hole size, we fabricated monolayer MoSe$_2$ flakes suspended over holes with a radius of 2.5, 5 and 7.5~$\upmu$m on gold-coated substrates (see Methods for fabrication details). Figure~\ref{fig:FIG3}\textbf{c} shows the temperature increase for these suspended monolayer flakes, where we normalize to the $\ln{(R/r_0)}$-factor, such that the slope is (inversely) proportional to $\kappa$ (see Eq.~(\ref{eq:1})). We find comparable results for holes with a radius of 5 and 7.5~$\upmu$m, while the sample with the 2.5-$\upmu$m hole gives rise to a lower $\Delta T$, and therefore a higher $\kappa_{\rm app}$. Raman thermometry measurements on crystals suspended over small holes are prone to several possible artefacts. It is possible that phonons with a relatively long mean free path are restricted by the size of the suspended region, which would lead to an underestimation of $\kappa$. For small hole sizes, any non-ideal heat sinking of the substrate-supported region of the crystal will also have a larger effect, because the smaller circumference implies a smaller region at the edge of the suspended crystal for heat sinking, leading to an underestimation of $\kappa$. Finally, laser absorption by the suspended crystal can be overestimated, as part of the incident light can be clipped by the small hole, as illustrated in Fig.~\ref{fig:FIG3}\textbf{d}, leading to an overestimation of $\kappa$. Since we find an apparent thermal conductivity for the 2.5-$\upmu$m hole that is more than two times larger than the intrinsic $\kappa$ we find using the larger holes, we attribute the artefact we observe for small holes to an error in determining $P_{\rm abs}$. A possible way to overcome this problem is to use a smaller laser spot size. However, this can introduce additional complications, such as non-diffusive heat transport~\cite{chiloyan2020nonfourier}, and does not resolve the other possible artefacts induced by small holes. We conclude that it is crucial to use gold-coated substrates with relatively large holes, with a radius of at least 5~$\upmu$m (see Fig.~\ref{fig:FIG3}\textbf{e}).

\subsection*{Experimental approach: effect of thickness}

\begin{figure*}[htbp]
\centering
\includegraphics[width=\linewidth]{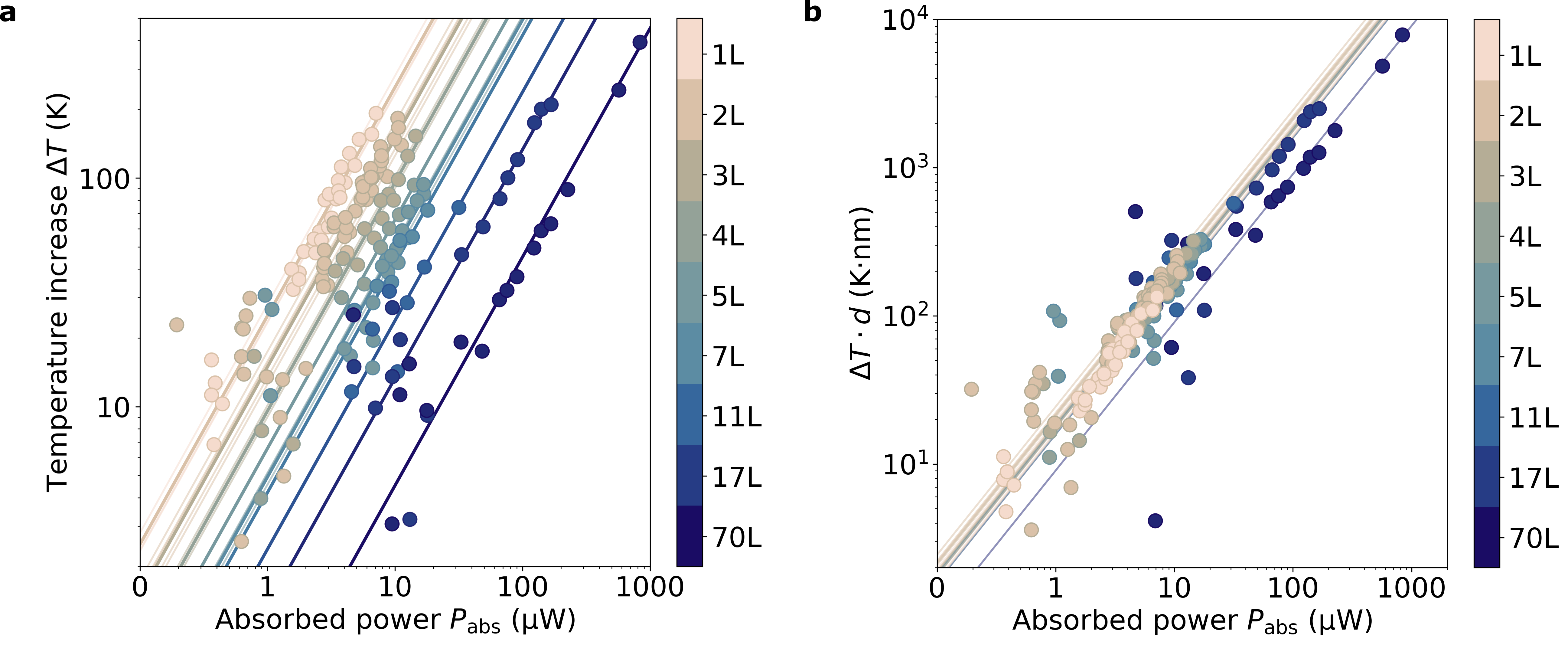}
\caption{\textbf{Raman thermometry of MoSe$_2$ as a function of crystal thickness.} \textbf{a)} Temperature rise $\Delta T$ as a function of absorbed power $P_{\rm abs}$ for MoSe$_2$ crystals of varying thickness.
\textbf{b)} The same data as in panel \textbf{a}, now multiplied by the thickness of each crystal, such that the slope is representative of $\kappa$. Each layer thickness has its own corresponding color (see color bars). Solid lines are linear fits to the data.}
\label{fig:FIG4}
\end{figure*}

Having identified and eliminated important artefacts by using gold-coated substrates with a hole radius of 7.5~$\upmu$m, as illustrated in Fig.~\ref{fig:FIG3}\textbf{e}, we proceed by studying the effect of crystal thickness on the thermal conductivity using our experimental approach of Raman thermometry. We perform these experiments on our artefact-free substrates and under vacuum conditions. We plot $\Delta T$ as a function of absorbed power (Fig.~\ref{fig:FIG4}\textbf{a}), and observe a clear trend with the thickness of the samples: thinner crystals heat up more significantly for the same absorbed power $P_{\rm abs}$. This is intuitive, as thinner crystals have a smaller volume in which the same amount of heat is deposited. Plotting $\Delta T \cdot d$ as a function of $P_{\rm abs}$ (see Fig.~\ref{fig:FIG4}\textbf{b}) gives a slope that is directly representative of the thermal conductivity $\kappa$ (see Eq.~(\ref{eq:1})). We now see that all data points fall on almost the same slope, suggesting that the intrinsic thermal properties of MoSe$_2$ are not dramatically affected by crystal thickness. A quantitative analysis of the experimental data using Eq.~(\ref{eq:1}) results in a $\kappa$ of $\sim$20~Wm$^{\shortminus1}$K$^{\shortminus1}$ for crystals with a thickness of 17L down to monolayer, and $\sim$40~Wm$^{\shortminus1}$K$^{\shortminus1}$ for bulk-like MoSe$_2$, with a thickness of $\sim$70L (Fig.~\ref{fig:FIG5}\textbf{a}). In the Supporting Fig.~\ref{fig:FigS-exp-lit}, we compare our values for the in-plane thermal conductivity with the available experimental results in the literature~\cite{zobeiri2019frequency,wang2018measurement,jiang2017probing,zhang2015measurement}. Most strikingly, our systematic thickness variation demonstrates a relatively weak effect of crystal thickness, within a factor two, whereas experimental literature values suggest a variation over almost an order of magnitude with particularly large conductivities for the thinnest crystals. We ascribe this discrepancy to the fact that not all measurements in the literature were performed under the same conditions nor with similar substrates, and often using non-coated substrates with rather small hole sizes, whereas in our case we used artefact-free substrates and constant fabrication and measurement conditions for all different thicknesses.

\subsection*{Theoretical approach: effect of thickness}

In order to interpret and understand our experimental results, we compute the thermal properties of MoSe$_2$ using density functional theory, as implemented in SIESTA~\cite{siesta}, in combination with the Temperature-Dependent Effective Potential (TDEP) method that allows to take into account phonons at a non-zero temperature~\cite{Hellman2011tdep1,Hellman2013tdep2}. In brief (see Methods for details), with this method we identify harmonic and anharmonic interatomic force constants taking into account atomic displacements and forces of a canonical ensemble at a given temperature. These computed force constants are the representation of the thermally averaged Born-Oppenheimer potential energy surface of the atomic displacements around the equilibrium positions. With this method we compute the phonon dispersion (see Supporting Fig.~\ref{fig:phononDispersion}) and the anharmonic terms of the interatomic potential, in order to obtain the in-plane lattice thermal conductivity $\kappa$. We compute $\kappa$ for bulk MoSe$_2$, and for 2D crystals with thicknesses from 6L down to the monolayer.

\begin{figure*}[htbp]
\centering
\includegraphics[width=15.5cm]{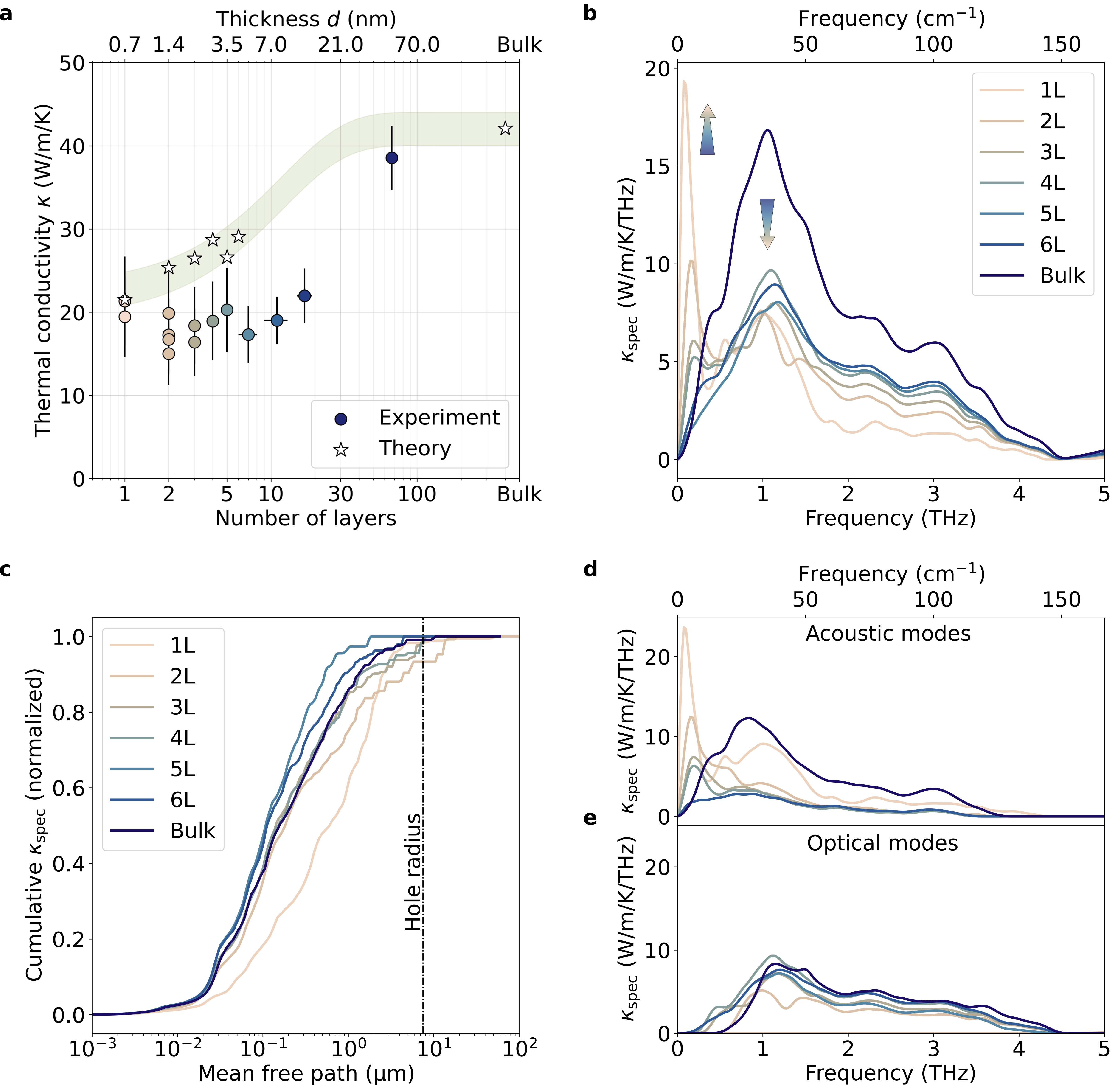}
\caption{\textbf{Microscopic understanding of heat transport in MoSe$_2$.} \textbf{a)}~In-plane thermal conductivity of MoSe$_2$ crystals as a function of thickness, using our experimental (circles) and theoretical (stars) approach. The shaded line is a guide to the eye for the theory results. \textbf{b)}~Spectrally decomposed thermal conductivity $\kappa_{\rm spec}$ as a function of phonon frequency,  indicating how towards thinner films an increasing contribution from a sub-THz mode compensates the decreasing contribution from modes around 1 THz. \textbf{c)}~Decomposed in-plane thermal conductivity as a function of phonon MFP. The cumulative thermal conductivity is normalized by the overall in-plane thermal conductivity. \textbf{d,e)}~ Spectrally decomposed thermal conductivity $\kappa_{\rm spec}$ for \textbf{(d)}~acoustic-like modes and \textbf{(e)}~optical-like modes.}
\label{fig:FIG5}
\end{figure*}

We compare the theoretically obtained in-plane thermal conductivity of MoSe$_2$ crystals with different thicknesses to the experimental results (see Fig.~\ref{fig:FIG5}\textbf{a}). For bulk MoSe$_2$, the experimental and theoretical values both give a $\kappa$ of $\sim$40~Wm$^{\shortminus1}$K$^{\shortminus1}$, which is close to the values found in the literature~\cite{Kandemir_2016}. For thinner crystals, we see that the first-principles-based results show a $\kappa$ of 21~Wm$^{\shortminus1}$K$^{\shortminus1}$ for monolayer MoSe$_2$, weakly increasing to 29~Wm$^{\shortminus1}$K$^{\shortminus1}$ for 6L. These values are in overall good agreement with the experiments, although the simulated values are slightly higher than the experimental results with $\sim$20~Wm$^{\shortminus1}$K$^{\shortminus1}$. 
Importantly, both results show that there is a weak effect of crystal thickness on the thermal conductivity. Furthermore, if there is any effect, it is opposite to the effect in graphite, which shows an increase in thermal conductivity upon decreasing crystal thickness, with monolayer graphene exhibiting the largest thermal conductivity~\cite{balandin2008superior,ghosh2010dimensional}. In Supporting Fig.~\ref{fig:theory_lit}, we compare our values for the in-plane thermal conductivity with the available results in the literature~\cite{kumar15,C5RA19747C,Gu2014dft,Kandemir_2016}. Moreover, we performed the same simulations for the TMD materials WSe$_2$ and MoS$_2$, which show a similar trend (see Supporting Fig.~\ref{fig:wse}). This suggests that the trend we observe both experimentally and theoretically is representative of the broader family of TMD materials.

Our simulation results provide important physical insights for the observed weak effect of crystal thickness on $\kappa$ for TMDs: we examine which phonons contribute to the total thermal conductivity by plotting the spectrally decomposed thermal conductivity of MoSe$_2$ $\kappa_{\rm spec}$ (see Fig.~\ref{fig:FIG5}\textbf{b}). We find that for bulk crystals, the largest contribution comes from modes around 1~THz. This contribution gradually decreases with the crystal thickness. However, towards the monolayer limit, modes with a frequency well below 1~THz start playing an important role. We confirm this picture by examining the phonon mean free path of each of the phonon modes in the decomposed thermal conductivity (see Supporting Fig.~\ref{fig:cum_kappa}). We show the cumulative thermal conductivity as a function of MFP (see Fig.~\ref{fig:FIG5}\textbf{c}), and observe that in the monolayer case, an increased fraction of the conductivity is carried by low-frequency modes with a relatively long MFP. This result also highlights the importance of using large hole sizes, as a significant fraction of $\kappa$ is carried by phonons with a MFP of several microns, which confirms that our experimental hole size is not significantly affecting the extracted $\kappa$ through edge scattering: phonons with a MFP $<$7.5~$\upmu$m contribute to $>$90\% of the total thermal conductivity.

In order to gain more understanding of the key phonon modes, we decompose the spectral contribution into acoustic modes (see Fig.~\ref{fig:FIG5}\textbf{d}) and low-frequency optical modes (see Fig.~\ref{fig:FIG5}\textbf{e}). For the latter, we only take into account of modes below 4~THz: the thermal conductivity of higher optical modes is negligible. The contribution of the optical modes, which are centered slightly above 1~THz and have an interlayer character, weakly decreases with decreasing crystal thickness. The acoustic contribution that is centered below 1~THz exhibits stronger thickness effects, with the most striking effect being the increasingly strong contribution of the flexural mode situated at $\sim$0.1~THz for thin MoSe$_2$. Thus, from the simulation results in Fig.~\ref{fig:FIG5}\textbf{b$-$e} we understand that towards the monolayer limit, the decreasing contribution to $\kappa$ from modes around 1~THz is compensated by the increasing contribution of modes with a much lower frequency, in particular a low-energy flexural mode, thus resulting in an overall weak effect of material thickness. 

This is a surprising result, because both the phonon dispersions and the phonon lifetimes (see Supporting Fig.~\ref{fig:lifetimes}) change drastically with thickness, as is also clear from the spectrally decomposed thermal conductivity in Fig.~\ref{fig:FIG5}. It is also surprising, because a strong effect of thickness was shown for graphene~\cite{balandin2008superior}. Moreover, it is remarkable that significant amounts of heat are carried by modes with a mean free path of several micrometers inside a material with sub-nanometer thickness. This shows that out-of-plane boundary scattering does not play any role for the in-plane thermal conductivity of 2D van der Waals bonded TMDs. This is in large contrast with thin films of 3D bonded materials, where the thermal conductivity is typically thought to be limited by boundary scattering at the film surface, limiting the mean free path out of plane to an effective scattering thickness. For 2D materials this is not the case: the very long lifetimes of low energy modes in thin MoSe$_2$ are made possible by the weakness of the van der Waals interlayer scattering, which is generic for all 2D materials, and leads to well known thermal transport anisotropy of more than an order of magnitude~\cite{jiang2017probing}. In our theoretical simulations, the full physical thickness is taken into account: surface vibrations are distinguished explicitly, and the scattering between bulk-localized and surface-localized modes is included in the anharmonic 3-phonon interatomic force constants. The simulated surface does not contain additional sources of scattering (strain, residues, defects, etc.) which would also limit the mean free path. The agreement with experiments is a further confirmation of the very clean and ideal nature of the experimental samples.

\begin{figure*}[htbp]
\centering
\includegraphics[width=\linewidth]{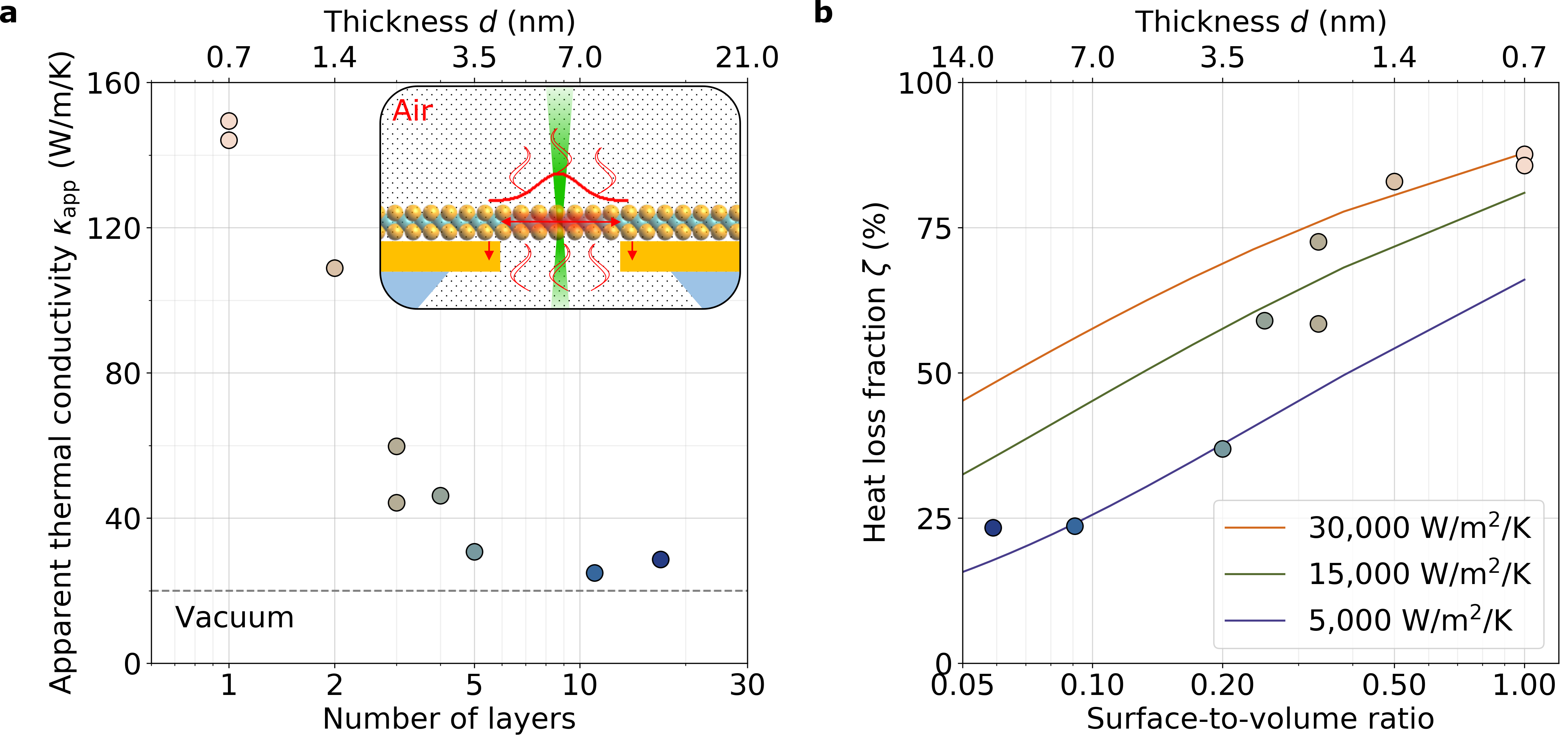}
\caption{\textbf{Air-mediated losses in suspended MoSe$_2$.} \textbf{a)}~Apparent in-plane thermal conductivity of suspended MoSe$_2$ flakes on large, gold-coated holes as measured in air. \textbf{b)}~Relative power losses to air, extracted by comparing measurements performed in vacuum with those performed in air. Solid lines represent the simulated power losses for different heat transfer coefficients (see Methods).}
\label{fig:FIG6}
\end{figure*}

\subsection*{Out-of-plane dissipation to the environment}

Many properties of thin, layered materials have been shown to be sensitive to the environment~\cite{gabourie2020reduced}. In the case of thermal properties, a relatively small effect caused by heat transport to gas molecules was observed for suspended graphene~\cite{chen2011raman}. We examine the effect of the surrounding environment on thermal transport in our MoSe$_2$ crystals, by performing Raman thermometry experiments both in vacuum and in air, for several samples with different thicknesses. In Fig.~\ref{fig:FIG6}\textbf{a}, we show the obtained apparent thermal conductivity $\kappa_{\rm app}$ as a function of flake thickness in the case of air, instead of vacuum. We find a thermal conductivity that is slightly higher in air than in vacuum for thick flakes, whereas it is almost an order of magnitude higher for monolayer MoSe$_2$. The reason for this large effect is likely that the presence of air introduces an additional cooling channel. In addition to in-plane diffusion from the hot spot to the heat sink, heat dissipation occurs by transfer to the ambient air as a sink (schematically depicted in the inset of Fig.~\ref{fig:FIG6}\textbf{a}). The relative effect of this competing dissipation channel is much larger than in the case of graphene, because the in-plane thermal conductivity of monolayer MoSe$_2$ is much smaller than that of graphene. We note that Eq.~(\ref{eq:1}) is not valid if there is an additional cooling channel, which means that the obtained apparent thermal conductivity $\kappa_{\rm app}$ in air is not an intrinsic material property of MoSe$_2$. However, it can be seen as an effective parameter describing heat transport in the combined air-MoSe$_2$ system. 

In order to understand the observed effect of the environment in more detail, we include additional cooling channels in our simulation of the Raman thermometry experiment (see Supporting Eq.~(\ref{eq:heat})). We first consider radiative cooling, estimating its maximum possible contribution by using a $\Delta T$ of 200~K, which is the largest value we used in our experiment (see Fig.~\ref{fig:FIG4}\textbf{a}). The results are shown in the Supporting Information, and indicate a negligible effect of $<$0.1\% for radiative cooling at such temperatures. Due to the $T^4$-scaling, this cooling channel will likely only start playing a role at significantly higher temperatures ($\Delta T \gg$ 200~K). The next cooling channel we consider is that of out-of-plane heat dissipation from MoSe$_2$ to the surrounding air molecules. We plot the experimentally obtained loss fraction, defined as $\zeta = 1 - \kappa_{\rm vac} / \kappa_{\rm app,air}$, as a function of surface-to-volume ratio of the crystal (see Fig.~\ref{fig:FIG6}\textbf{b}), and compare it to our simulation of the Raman thermometry experiment that includes an out-of-plane heat transfer term. We find a heat loss fraction $\zeta$ of $\sim$20\% for the lowest surface-to-volume ratio, which we can reproduce with a heat transfer coefficient $h_{\rm c}$ of $\sim$5,000~Wm$^{\shortminus2}$K{$^{\shortminus1}$}. For monolayer MoSe$_2$, on the other hand, we find $>$80\% loss, which we can reproduce with a heat transfer coefficient $h_{\rm c}$ of $\sim$30,000~Wm$^{\shortminus2}$K{$^{\shortminus1}$}. These are much larger values than the typical values for the convective heat transfer coefficient found in the literature~\cite{zhang2015measurement,easy2021experimental}, even for forced convection by gases: $h_{\rm c} = 25-250$~Wm$^{\shortminus2}$K{$^{\shortminus1}$}~\cite{BergmanHeatTransfer}. Our value, however, is very close to the value observed for monolayer graphene (2.9$\times$10$^4$ Wm$^{\shortminus2}$K{$^{\shortminus1}$}~\cite{chen2011raman}), and close to the ideal heat transfer coefficient to air at ambient pressure and temperature with an ideal molecular accommodation coefficient (10$^5$ Wm$^{\shortminus2}$K{$^{\shortminus1}$}~\cite{chen2011raman}). Importantly, these results provide clear evidence that out-of-plane heat dissipation to air plays an important role in the cooling dynamics of suspended ultrathin materials, and that cooling is significantly more efficient for atomically thin crystals than for thicker crystals. Importantly, when such thin crystals are placed in air, their overall cooling ability is enhanced by their efficient interaction with air molecules. This is very relevant and beneficial for designing applications where the thermal management of TMDs and other layered materials is a crucial consideration. 

\section{Conclusion}

We used Raman thermometry and \textit{ab initio} simulations to investigate the influence of thickness on the thermal conductivity of suspended MoSe$_2$ crystals. We observed excellent agreement between our experimentally measured and computed in-plane lattice conductivities. Both approaches indicate a relatively weak effect of crystal thickness on the lattice thermal conductivity $\kappa$ -- within a factor two. We explain this weak thickness influence as the result of competing effects in the phonon contribution to the thermal conductivity. Furthermore, we have shown the importance of a careful experimental configuration in order to obtain reliable thermal conductivities from Raman thermometry measurements. Finally, we have demonstrated a very strong effect of the environment on thermal transport, in particular in the case of monolayer MoSe$_2$, which is caused by out-of-plane heat dissipation with a surprisingly large heat transfer coefficient. We note that many of these results represent essential guidance for the thermal investigation of other TMD materials. This work provides a basis to understand and engineer thermal transport properties of a broad class of materials, with promising applications in thermal management, energy materials and (opto)electronic devices.  

\section{Methods}
\textbf{Sample fabrication}. The sample fabrication, based on PDMS-assisted dry transfer of mechanically exfoliated MoSe$_2$ flakes (HQ graphene, 2H phase), is described in detail in Ref.~\cite{Varghese2021b}. As substrates, we used holey Si$_3$N$_4$ membranes (Norcada, NTPR005D-C15) for the study of the effect of crystal thickness and the effect of gold coating, (see samples in Fig.~\ref{fig:FIG1}\textbf{b} and Fig.~\ref{fig:FIG1}\textbf{d}). Those substrates have a single hole with a radius of 7.5~$\upmu$m. For the study of the effect of hole size, we used gold-coated silicon-on-insulator wafers with back-thinned membranes. We used focused ion beam to perforate holes with a radius of 2.5 and 5~$\upmu$m (see Fig.~\ref{fig:FIG1}\textbf{c}). The gold coatings, consisting of 50~nm gold with 5~nm titanium adhesion layer, were deposited prior to transfer using E-beam evaporation (AJA Orion).

\textbf{Raman thermometry}. Raman spectra were collected with a Horiba T64000 Raman spectrometer and a laser beam, with a wavelength of $\lambda = 532$~nm, focused to a $1/e$ spot size of $\sim$1~$\upmu$m (see Supporting Fig.~\ref{fig:spotsize} for spot size measurements). For thermal measurements, the samples were placed in a temperature controlled vacuum stage (Linkam). The samples were glued onto a holey Cu plate using silver paste, for a good thermal link with the stage. The samples were left to thermalize for 20~min at each temperature. These calibration measurements were taken both in the supported and suspended regions, giving comparable results (see Supporting Information). The temperature increase is defined as $\Delta T = (\nu_P - \nu_{P=0}) / \chi_T$, with $\nu_{P=0}$ the intercept from the linear fit of Raman shift with laser power. The Raman experiments were performed both in vacuum ($5 \cdot 10^{\shortminus3}$~mbar) and air ($1$~bar). The absorbance of each suspended MoSe$_2$ crystal was determined using a home-built optical setup by measuring transmittance and reflectance through the suspended region, see Supporting Information. We note that the experimental $\kappa$ of the bulk material is extracted using an effective thickness $d_{\rm eff}$ equal to the penetration depth in MoSe$_2$ ($d_{\rm eff} = \lambda /4\pi k = 20$~nm, with the extinction coefficient $k = 2.08$ at 532~nm~\cite{beal1979kramers}) as the thickness of the flake, instead of the actual thickness of 47~nm, because of the very low out-of-plane thermal conductivity, which means that only a region with a thickness $d_{\rm eff}$ carries the in-plane heat. For other thicknesses, we assume homogeneous heating in the $c$-axis of the flake. 

\textbf{Density Functional Theory (DFT) simulations}. Our computational approach is based on first-principles calculations. We study thermal transport properties using the density functional theory as implemented in the SIESTA program~\cite{Soler_2002, siesta} and employing LMKLL functionals~\cite{LMKLL} to take into account van der Waals interactions. We consider structures with a different number of layers, from monolayer up to 6L, with 17~\AA~of vacuum to eliminate the interaction between periodically repeated images.
Calculations are converged with 1000~Ry energy cutoff for the real-space grid with a ($20 \times 20 \times 1$) \textbf{k}-points sampling of the Brillouin zone for all the layers and ($20 \times 20 \times 20$) \textbf{k}-points grid for the bulk. A standard double zeta polarized  (DZP) basis  for Mo and Se atoms and an electronic temperature of 300~K was used. The conjugate gradient algorithm is used to relax the cell and the atomic positions until the forces on the atoms became smaller than 0.001~eV/\AA~and the maximum stress component is smaller than 0.5~GPa.
The calculations of forces and stress were then performed with ($10 \times 10 \times 1$) supercells and ($8 \times 8 \times 2$) supercells for the bulk material with the standard diagonalization method. The number of atoms in the supercells varies from 192 atoms in the monolayer to 1152 atoms in the 6-layer flake. The thermal properties are then computed with the TDEP method. The convergence of forces in TDEP required 7 iterations, where an iteration consists in generating a set of displacements, computing forces and fitting force-constants. The temperature used to generate snapshots is 300~K. To better average the forces, the number of configurations used in the procedure was increased as a geometrical series, with the 7th iteration computed using 128 configurations. The thermal conductivity is calculated by iteratively solving the full Boltzmann transport equation (BTE) for several \textbf{q}-point grid densities and extrapolating the value for an infinite number of \textbf{q}-points.

\section*{Acknowledgements}

The authors thank Andrea Pitillas Martínez for the graphics shown in the TOC and Figure~\ref{fig:FIG1}\textbf{a} and \ref{fig:FIG1}\textbf{b}. 
D.S.R. and S.V. would like to acknowledge the support of the Spanish Ministry of Economy through FPI-SO2019 and FPI-SO2018, respectively. 
R.F., P.O. and Z.Z. acknowledge support by the EU H2020-NMBP-TO-IND-2018 project ``INTERSECT'' (Grant No. 814487), the EC H2020-INFRAEDI-2018-2020 MaX ``Materials Design at the Exascale'' CoE (Grant No. 824143), and Spanish MCI/AEI/FEDER-UE (Grant No. PGC2018-096955-B-C43). 
O.H. acknowledges support from the Swedish Research Council (VR) program 2020-04630. 
P.W. acknowledges funding from the European Union's Horizon 2020 research and innovation program under the Marie Skłodowska-Curie Grant Agreement No. 754510 (PROBIST). 
M.S., A.E.S., E.C.A. and C.M.S.T. acknowledge support of the Spanish MICIN project SIP (PGC2018-101743-B-I00). 
S.O.V. acknowledges support from MINECO under contract numbers PID2019-111773RB-I00/AEI/10.13039/501100011033. 
Z.Z. acknowledges financial support by the Netherlands Sector Plan program 2019-2023. 
M.J.V. acknowledges support from FRS-FNRS Belgium PdR Grant No. T.0103.19 - ALPS, and contributions from the Melodica flag-era.net project. 
K.J.T., M.S., C.M.S.T., S.O.V. and N.F.v.H. acknowledge funding from BIST Ignite project 2DNanoHeat. 
K.J.T. acknowledges funding from the European Union's Horizon 2020 research and innovation program under Grant Agreement No. 804349 (ERC StG CUHL), RYC fellowship No. RYC-2017-22330, and IAE project PID2019-111673GB-I00. 
ICN2 was supported by the Severo Ochoa program from Spanish MINECO Grant No. SEV-2017-0706 and Generalitat de Catalunya (CERCA program and Grant 201756R1506).



\bibliography{references.bib}


\clearpage

\newpage
\onecolumngrid

\SupplementaryMaterials

\section*{Supporting Information}

\subsection{Simulation of Raman thermometry experiment} 

We simulate thermal transport in the geometry corresponding to our Raman thermometry experiment by solving the 2D (in-plane) steady state thermal model, \textit{i.e.}\ Fourier's law, including a heat exchange at the thin sample's surfaces, \textit{e.g.}\ due to convection and/or conduction to the environment, as well as thermal radiation. In polar coordinates with radial symmetry, the time-dependent temperature rise above ambient temperature $\Delta T(r,t)$ evolves according to~\cite{chen2011raman}
\begin{equation} \label{eq:heat}
    \frac{\partial \Delta T}{\partial t} = 
        \underbrace{\frac{1}{r} \cdot \frac{\partial}{\partial r} \left ( r \cdot \frac{\kappa}{C_v} \cdot \frac{\partial \Delta T}{\partial r} \right )}_\text{in-plane conduction}
    + \underbrace{\frac{S}{C_v}}_\text{source} 
    - \underbrace{\frac{N_\mathrm{surf} \cdot h_c \cdot \Delta T}{d \cdot C_v}}_\text{out-of-plane dissipation} 
    - \underbrace{\frac{N_\mathrm{surf} \cdot \sigma_\mathrm{SB} \cdot \Delta T^4}{d \cdot C_v}}_\text{radiation} \hspace{1cm}. \end{equation}
Here, the MoSe$_2$ sample's volumetric heat capacity is $C_v = 1.87$~MJ/m$^3$/K (the bulk value)~\cite{C5RA19747C}, $\kappa$ is its in-plane thermal conductivity and $d$ its thickness. We assume perfect heat sink conditions $\Delta T(R,t) = 0$ at the hole boundaries, with $R = 7.5$~$\upmu$m.

The out-of-plane heat transfer coefficient $h_{\rm c}$ is set to $0$ to simulate the experiment in vacuum, but then adjusted to simulate the experiment in air. Similarly, we exclude the radiative term, with the Stefan-Boltzmann constant $\sigma_\mathrm{SB} = 5.67 \cdot 10^{-8}$W/m$^2$/K$^4$, except in  Supp.~Fig.~\ref{fig:Alex1}\textbf{b}. $N_\mathrm{surf} = 2$ is the number of surfaces (top and bottom).

The Gaussian volumetric heating source $S(r)$ is related to the total absorbed heating laser power $P_\mathrm{abs}$ and the 1/e spot size $r_0$ via~\cite{cai2010thermal}
\begin{equation} \label{eq:source}
S(r) = \frac{P_\mathrm{abs}}{d \pi r_0^2} \cdot \exp \left (-\frac{r^2}{r_0^2} \right ) \hspace{1cm}.
\end{equation}
The steady state $(\partial \Delta T / \partial t = 0)$ is found by evolving Eq.~(\ref{eq:heat}) in the time domain, starting from $\Delta T(r,0) = 0$, by the forward time centered space (FTCS) method until a steady state $\Delta T(r, \infty)$ is reached, typically after 10~$\upmu$s. We take care to fulfill the von Neumann stability condition for FTCS in 2D, $(\kappa/C_v) \Delta t/\Delta r^2 < 0.25$.

The comparison with the experimental situation is achieved by extracting the part of the steady state heat profile that is being probed by the Raman laser. This is done by taking the average of the radial profile $\Delta T(r, \infty)$ within the laser spot area, \textit{i.e.} from $r=0$ to $r=r_0$. This final $\Delta T_\mathrm{avg}$ is used to obtain the calculated heating rate $\partial T / \partial P_\mathrm{abs}$. We then compare this calculated $\partial T / \partial P_\mathrm{abs}$ with the one we obtain from Eq.~(\ref{eq:1}), using $\alpha = 1$, for a given thermal conductivity $\kappa$, and three different crystal thicknesses $d$ (see Fig.~\ref{fig:Alex1}\textbf{a}). We find excellent agreement -- within 2\% -- between our full Raman thermometry simulation and the simplified Raman thermometry equation (Eq.~(\ref{eq:1}) in the main text), see Supp.~Fig.~\ref{fig:Alex1}\textbf{a}, which confirms the validity of both approaches. 

\clearpage

\subsection{Exfoliated \texorpdfstring{MoSe$_2$} ~~flakes on PDMS and thickness determination}
\begin{figure}[ht!]
    \centering
    \includegraphics[width=\linewidth]{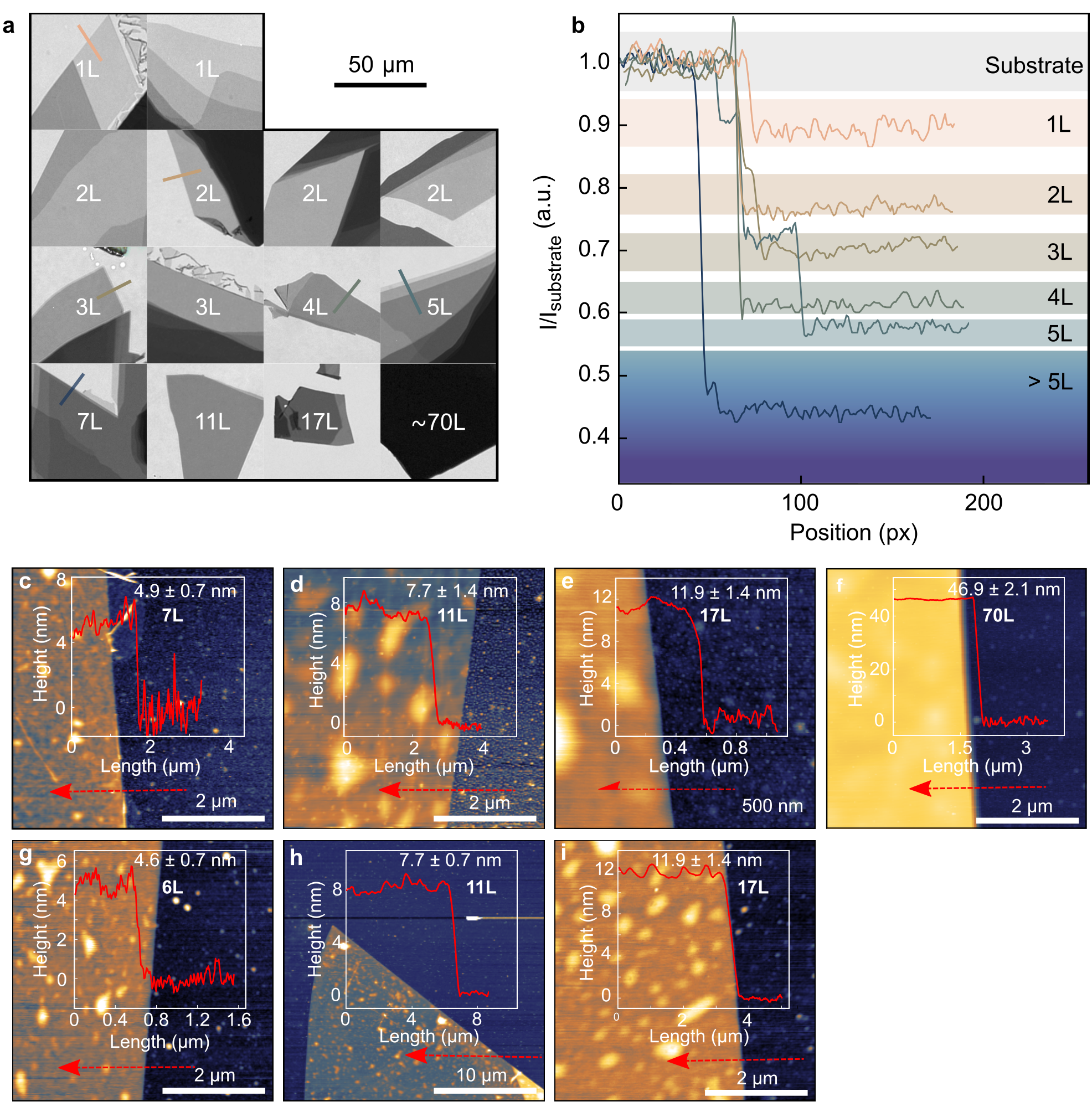}
    \caption{\textbf{Exfoliated MoSe$_2$ flakes and thickness determination. a)}~Green channel of transmission optical images for MoSe$_2$ flakes exfoliated on PDMS. Solid lines correspond to the intensity profiles across the lines marked in \textbf{(b)}~normalized to the substrate intensity. \textbf{(c$-$i)}~Atomic force microscope images of MoSe$_2$ flakes transferred onto \textbf{(c$-$f)}~gold-coated substrates and \textbf{(g$-$i)}~Si$_3$N$_4$ substrates. Insets show exemplary height profiles across the edge of the MoSe$_2$ flakes. The optical contrast has been calibrated using photoluminescence measurements, see Ref.~\cite{babacic2021thickness}}.
    \label{fig:FIGS1}
\end{figure}

\newpage
\subsection{Raman spectra as a function of incident laser power on suspended \texorpdfstring{MoSe$_2$} ~~crystals}
\begin{figure}[ht!]
    \centering
    \includegraphics[width=\linewidth]{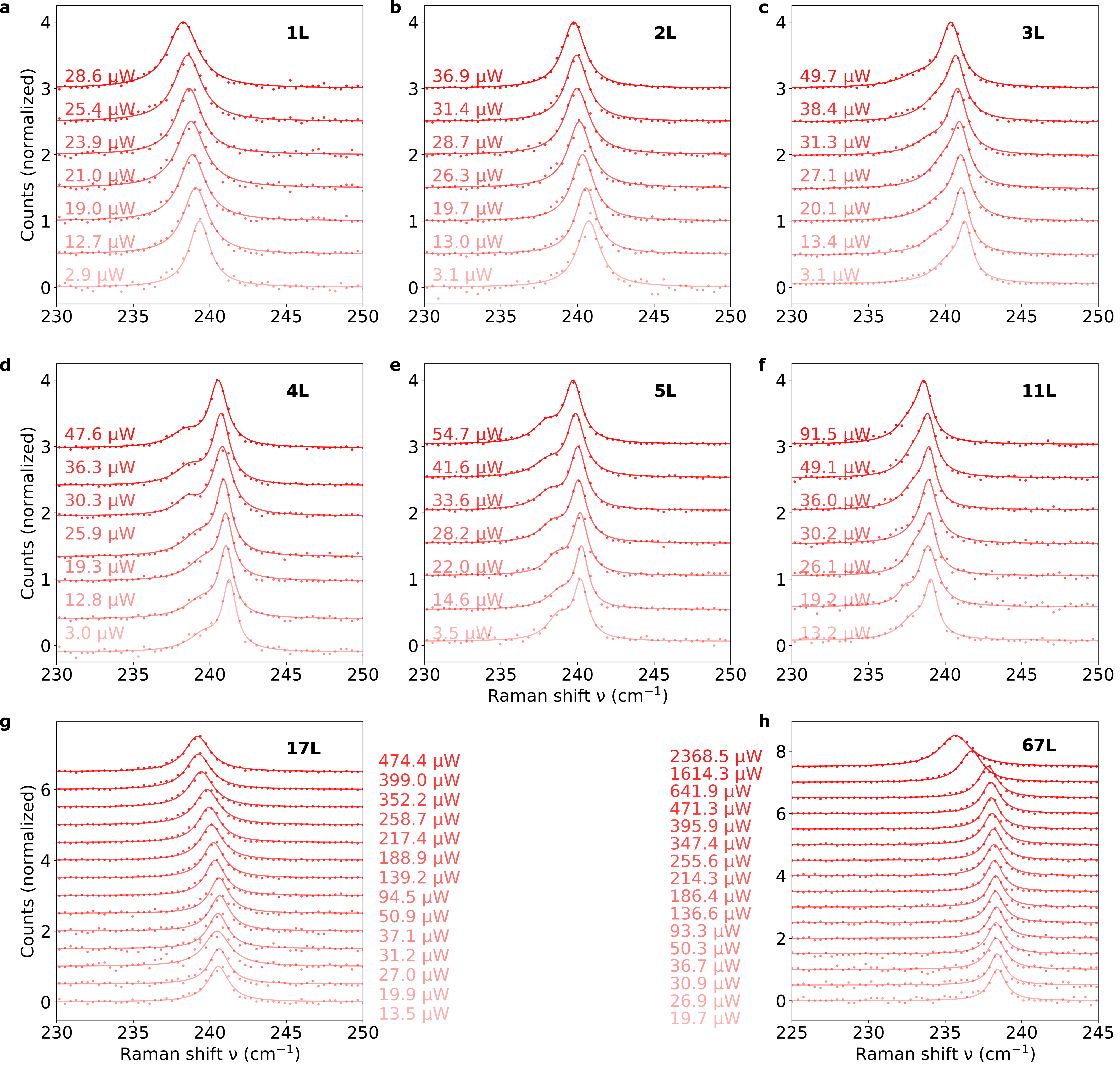}
    \caption{\textbf{Raman thermometry on suspended MoSe$_2$ crystals.} Raman spectra in vacuum as a function of laser power at sample position, for suspended MoSe$_2$ flakes with thickness ranging from 1L to $\sim$70L (dots), with corresponding Lorentzian fits (lines). The flakes are suspended on gold-coated substrates with holes with a radius of 7.5~$\upmu$m. Raman spectra of MoSe$_2$ flakes $>$3L were fitted with two Lorentzian resonances. From these measurements we obtain the heating rate $\partial T / \partial P_{\rm abs}$, from which we determine the thermal conductivity $\kappa$. }
    \label{fig:FIG-rawP_vacuum}
\end{figure}

\newpage
\subsection{Temperature calibrations at suspended and supported regions of \texorpdfstring{MoSe$_2$} ~~crystals}
\begin{figure}[ht!]
    \centering
    \includegraphics[width=\linewidth]{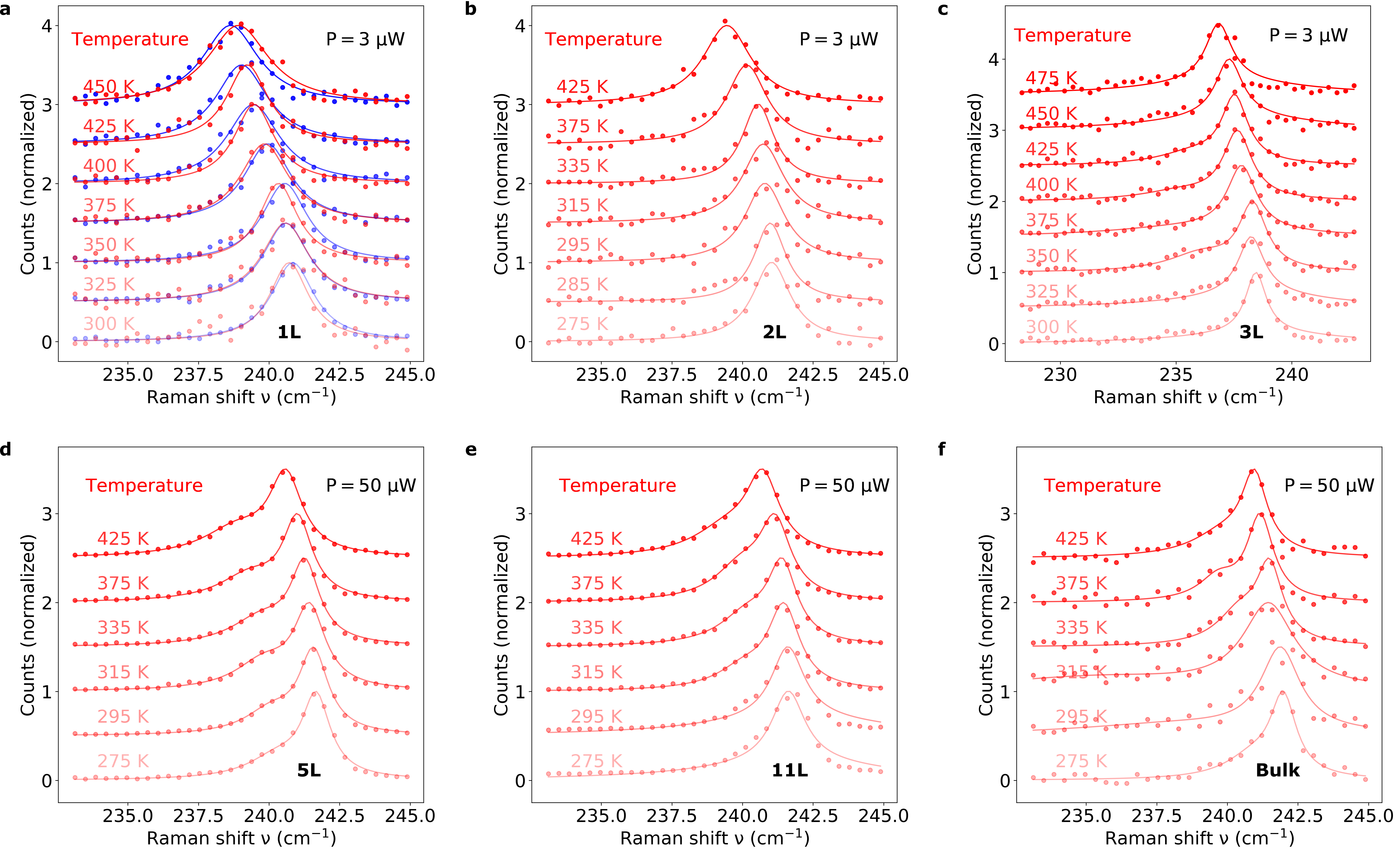}
    \caption{\textbf{Temperature calibrations.} \textbf{a$-$f)} Raman spectra at different sample stage temperature taken at the \textbf{(a$-$c)} suspended regions for 1L, 2L and 3L, and \textbf{(d$-$f)} supported regions for 5L, 11L and bulk. Blue data in panel \textbf{(a)} correspond to measurements on the supported region, indicating that the red-shift is very similar for supported and suspended regions. From these measurements we determine the temperature coefficients $\chi_{\rm T} = \partial \nu / \partial T$, which we use to convert power-dependent Raman shifts to a power-dependent temperature increase. }
    \label{fig:temp_calib}
\end{figure}

\newpage
\subsection{Thermal transport simulation}
\begin{figure}[ht!]
    \centering
    \includegraphics[width=\linewidth]{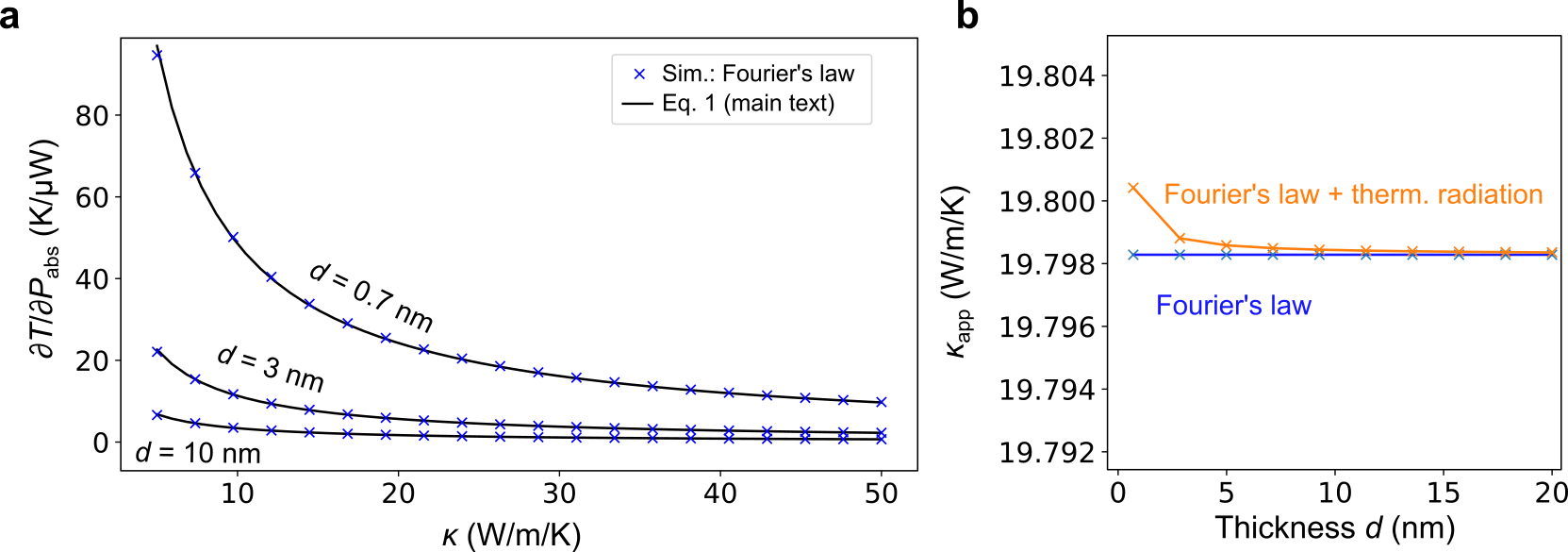}
    \caption{\textbf{Thermal transport simulation.} \textbf{a)}~Simulated thermal power response as a function of input thermal conductivity $\kappa$ (crosses), together with the prediction from Eq.~(\ref{eq:1}) in the main text (solid line) using $\alpha = 1$. Convective and radiative heat transfer contributions are set to zero here. \textbf{b)}~Effect of thermal radiation, for a maximum heating $\Delta T$ of 200~K at each thickness. We find a maximum difference of only 0.1\% between including and excluding the thermal radiation.}
    \label{fig:Alex1}
\end{figure}

\newpage
\subsection{Effect of the heat sink on thermal conductivity}
\begin{figure}[ht!]
    \centering
    \includegraphics[width=\linewidth]{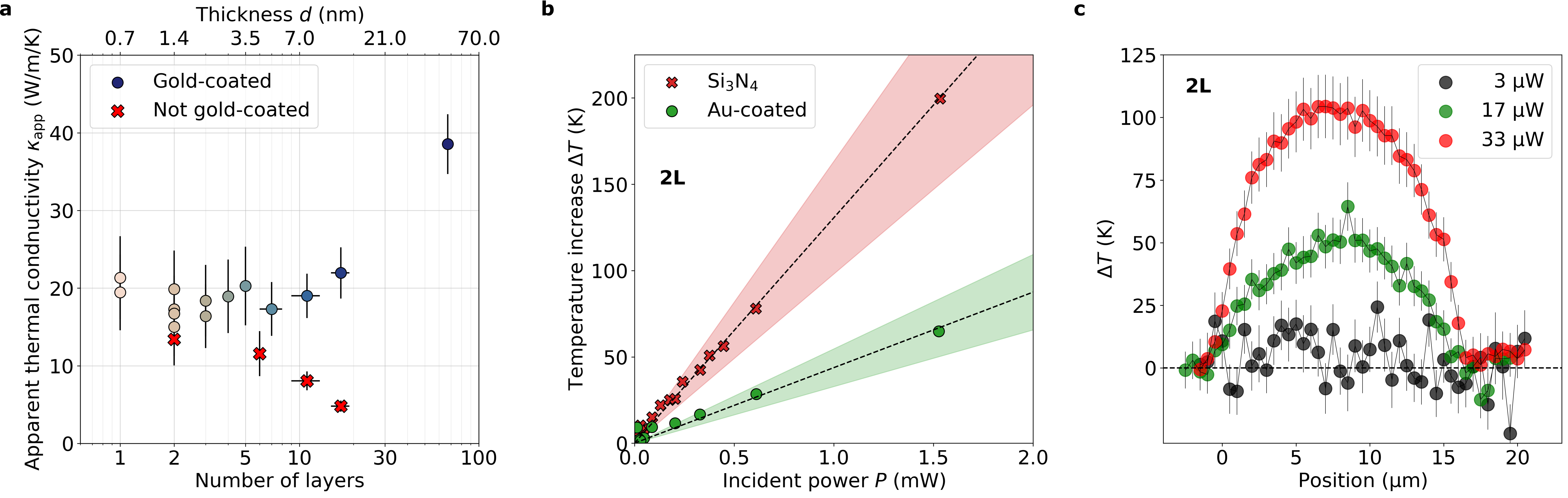}
    \caption{\textbf{Effect of the heat sink.} \textbf{a)}~Apparent in-plane thermal conductivity of suspended MoSe$_2$ flakes on Si$_3$N$_4$ (red crosses), compared to $\kappa$ obtained for those suspended on gold-coated substrates (circles).
    \textbf{b)}~Comparison between temperature increase on the supported regions of two bilayer MoSe$_2$ flakes with (green circles) and without (red crosses) gold coating. Flakes supported on gold-coated substrates heat up more than 3$\times$ less than those on Si$_3$N$_4$ substrates. \textbf{c)}~Temperature increase measured across the center of the suspended region of a bilayer flake on a gold-coated substrate, using one-laser scanning Raman thermometry. The suspended region heats up significantly with an incident power of a few tens of $\upmu$W, whereas the supported region does not. This shows that heat sinking is efficient for this range of incident power, which corresponds to the power used in the main Raman thermometry experiments used to determine $\kappa$.}
    \label{fig:FigS-coating}
\end{figure}

\newpage
\subsection{Experiment vs literature}
\begin{figure}[ht!]
    \centering
    \includegraphics[width=0.5\linewidth]{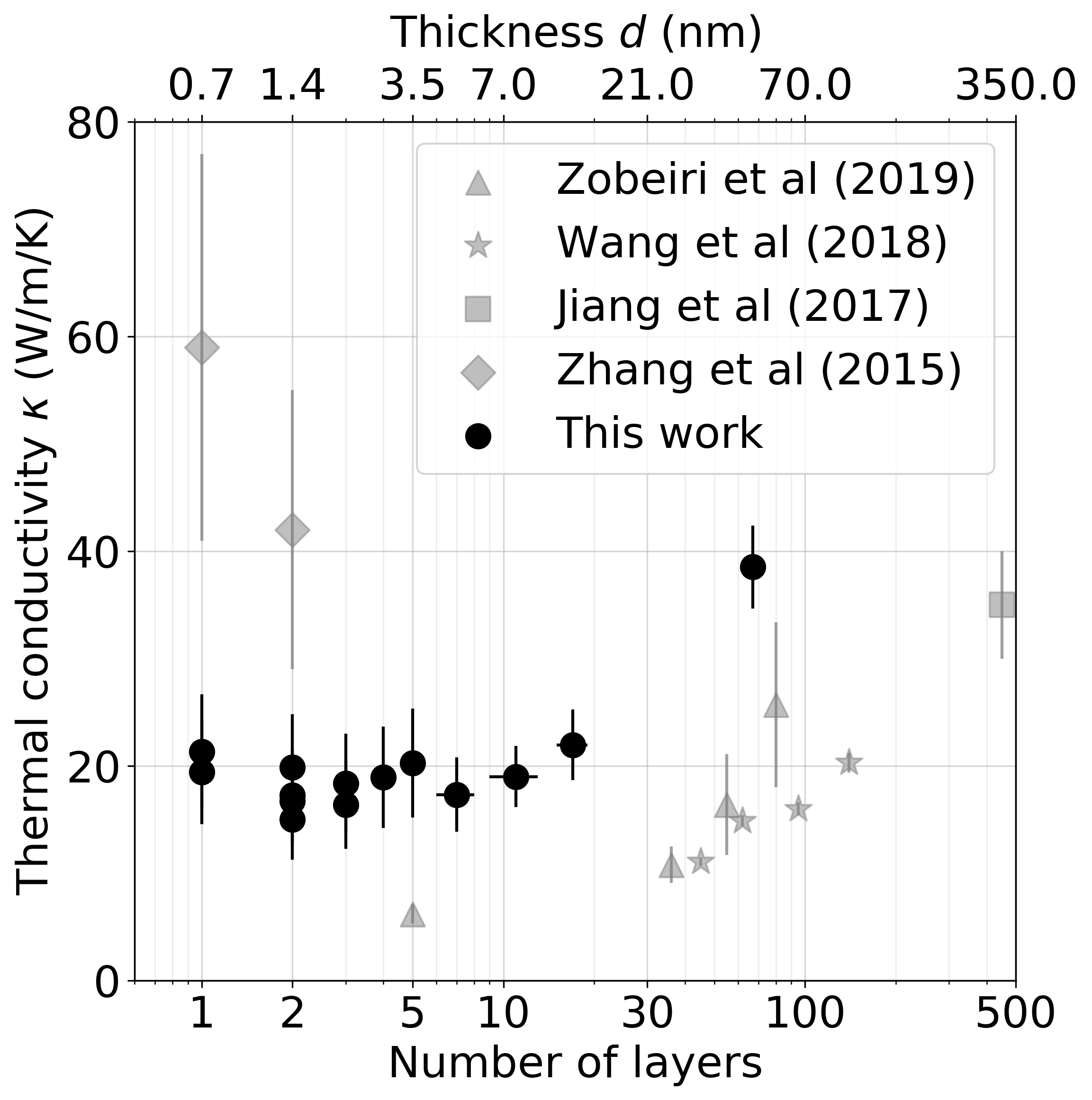}
    \caption{\textbf{Experiment vs literature.} Experimental in-plane thermal conductivity of MoSe$_2$ obtained in this work (black) and that reported in the literature (gray) by Zobeiri \textit{et al.}~\cite{zobeiri2019frequency}, Wang \textit{et al.}~\cite{wang2018measurement}, Jiang \textit{et al.}~\cite{jiang2017probing} and Zhang \textit{et al.}~\cite{zhang2015measurement}, all using Raman thermometry.}
    \label{fig:FigS-exp-lit}
\end{figure}

\newpage
\subsection{Phonon dispersions for different \texorpdfstring{MoSe$_2$} ~~layers}
\begin{figure}[ht!]
    \centering
    \includegraphics[width=\linewidth]{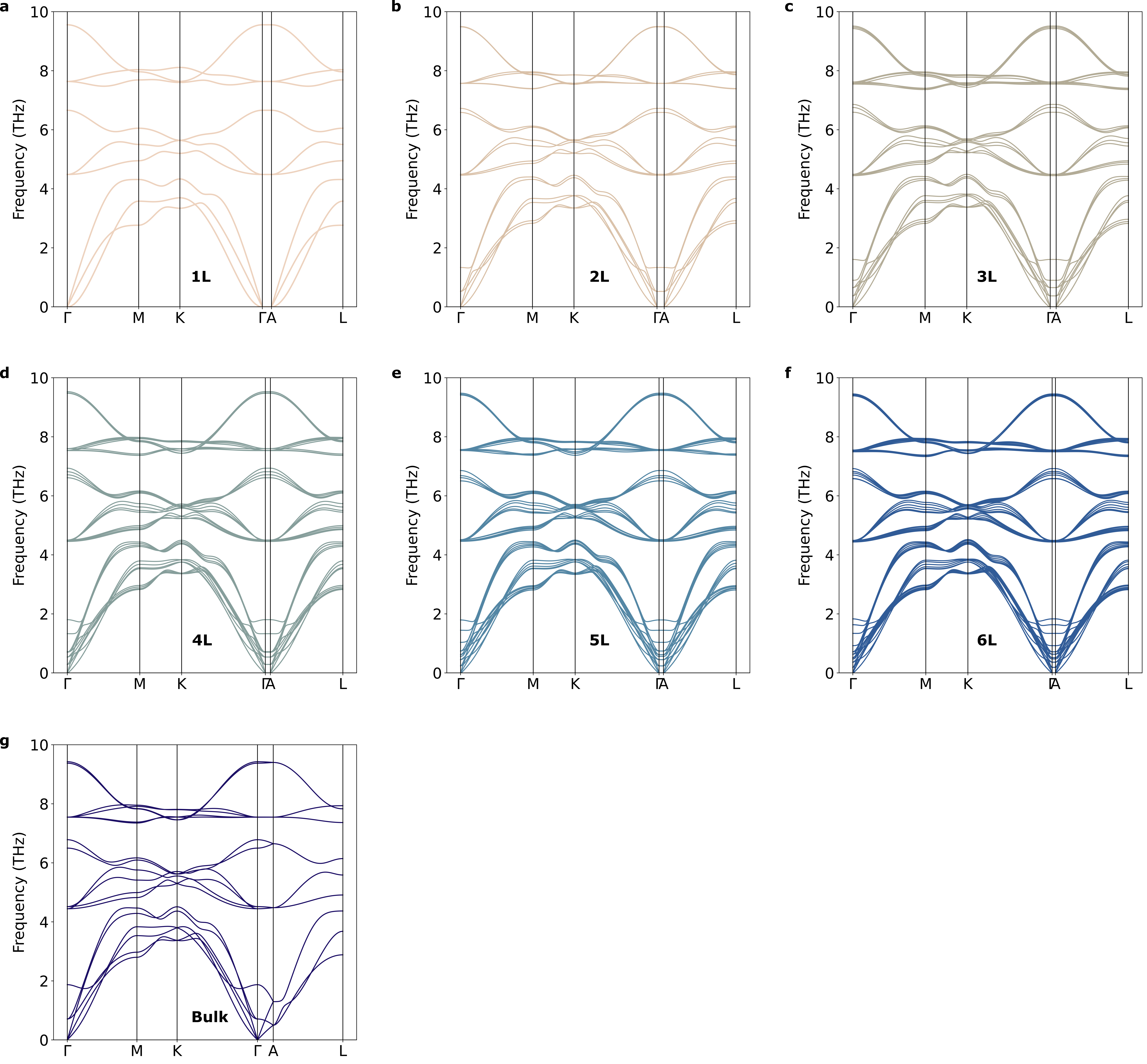}
    \caption{\textbf{Calculated phonon dispersions for different MoSe$_2$ layers.}  \textbf{a)} Monolayer, \textbf{(b)}~bilayer, \textbf{(c)} 3 layers, \textbf{(d)} 4 layers. \textbf{(e)} 5 layers, \textbf{(f)} 6 layers and \textbf{(g)} bulk structure. }
    \label{fig:phononDispersion}
\end{figure}

\newpage
\subsection{Theory vs literature}
\begin{figure}[ht!]
    \centering
    \includegraphics[width=0.5\linewidth]{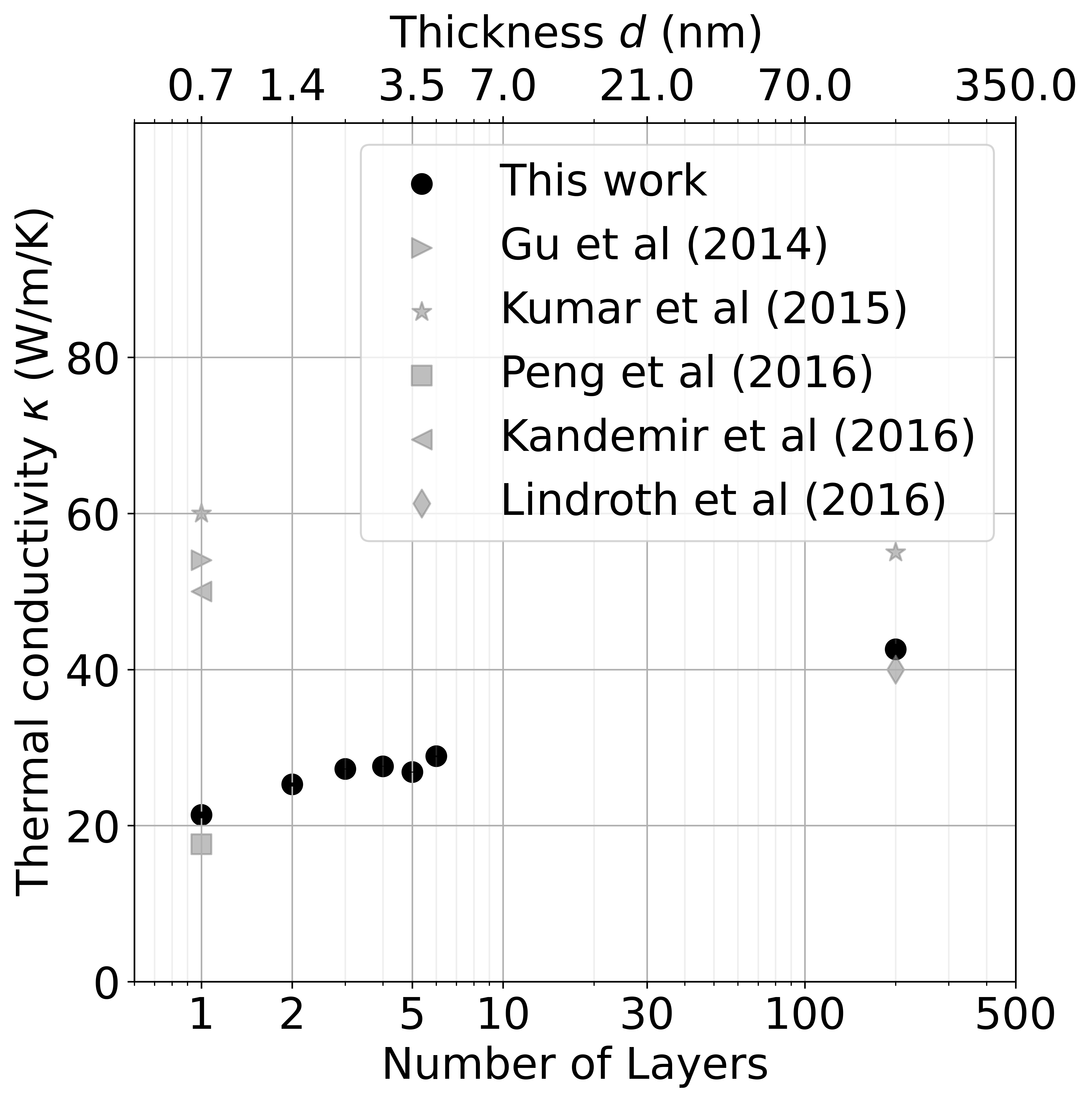}
    \caption{\textbf{Theory vs literature.} Computed in-plane thermal conductivity of MoSe$_2$ obtained in this work (black) and reported in the literature (gray) by Peng \textit{et al.}~\cite{C5RA19747C}, Gu \textit{et al.}~\cite{Gu2014dft}, Kumar \textit{et al.}~\cite{kumar15}, Kandemir \textit{et al.}~\cite{Kandemir_2016} and Lindroth  \textit{et al.}~\cite{lindorth}. We note that the difference between our theoretical prediction of $\kappa$ and previous theoretical works can be attributed to the different approaches and accuracy used. Previous works reported lattice thermal conductivity predicted with MD simulations which strongly depend on the force-field used in the calculations. Another difference can be found in the \textbf{q}-points grid used for the calculations. We extrapolated $\kappa$ by fitting the values obtained for different grids of \textbf{q}-points, since a finite \textbf{q}-point grid usually underestimates the results. The values found with this method are in good agreement with our experiment and the small differences can be attributed to the approximations that both experiment and theory use to determine the thermal conductivity.}
    \label{fig:theory_lit}
\end{figure}

\newpage
\subsection{Calculated in-plane thermal conductivity of \texorpdfstring{MoSe$_2$, WSe$_2$ and MoS$_2$} ~~} 
\begin{figure}[ht!]
    \centering
    \includegraphics[width=0.5\linewidth]{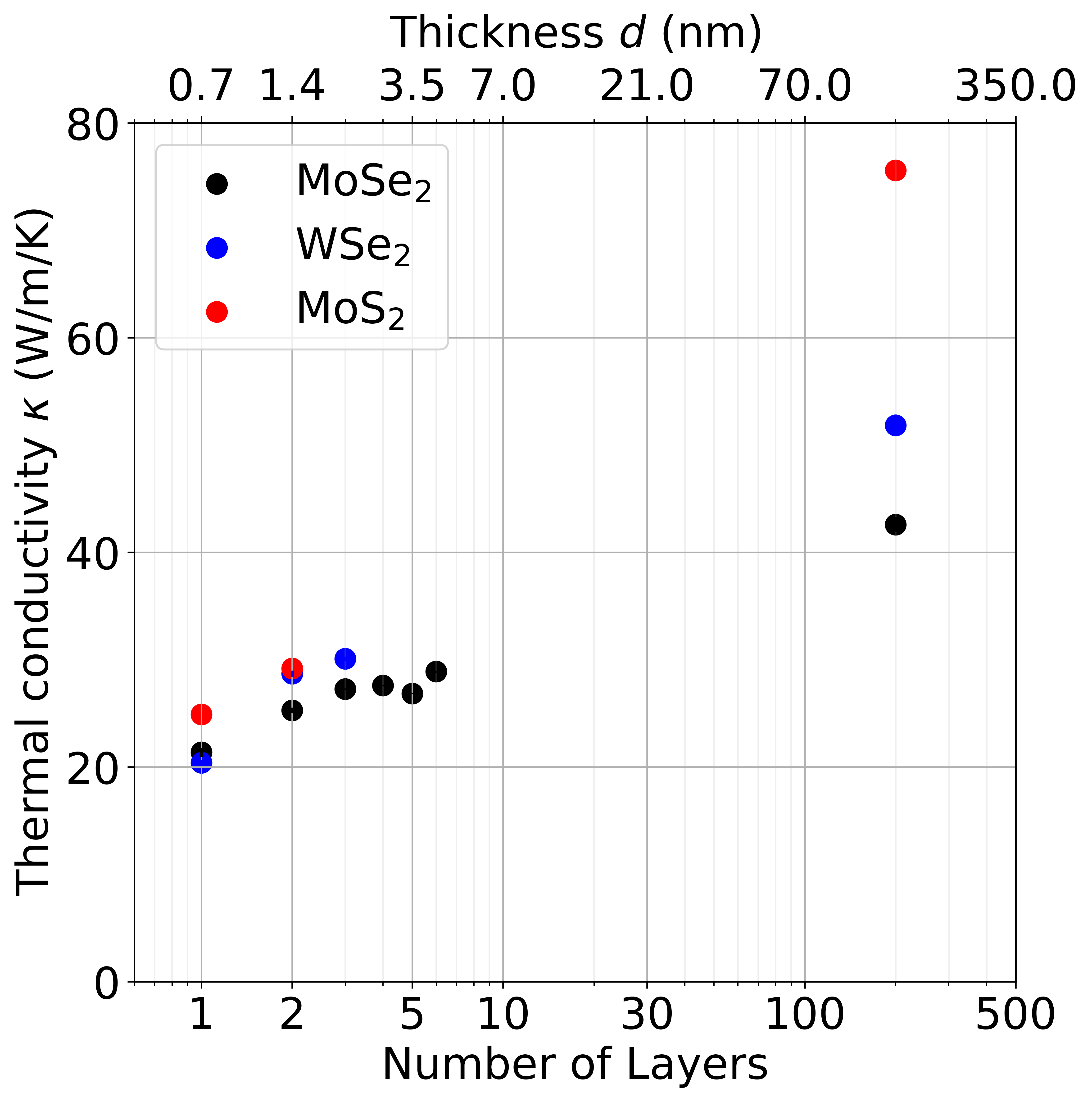}
    \caption{\textbf{MoSe$_2$ vs WSe$_2$ and MoS$_2$ theory.} Computed in-plane thermal conductivity of MoSe$_2$ (black), WSe$_2$ (blue) and MoS$_2$ (red). The small difference in $\kappa$ is mainly related to the difference in the atomic mass between W, Mo, S and Se. Importantly, all the materials show the same trend with crystal thickness, which is a weak increase with thickness. This suggests that the results we discuss for MoSe$_2$ are generally representative for the family of layered TMD materials.}
    \label{fig:wse}
\end{figure}

\newpage
\subsection{Phonon lifetimes for monolayer and bulk \texorpdfstring{MoSe$_2$} ~~}
\begin{figure}[ht!]
    \centering
    \includegraphics[width=0.5\linewidth]{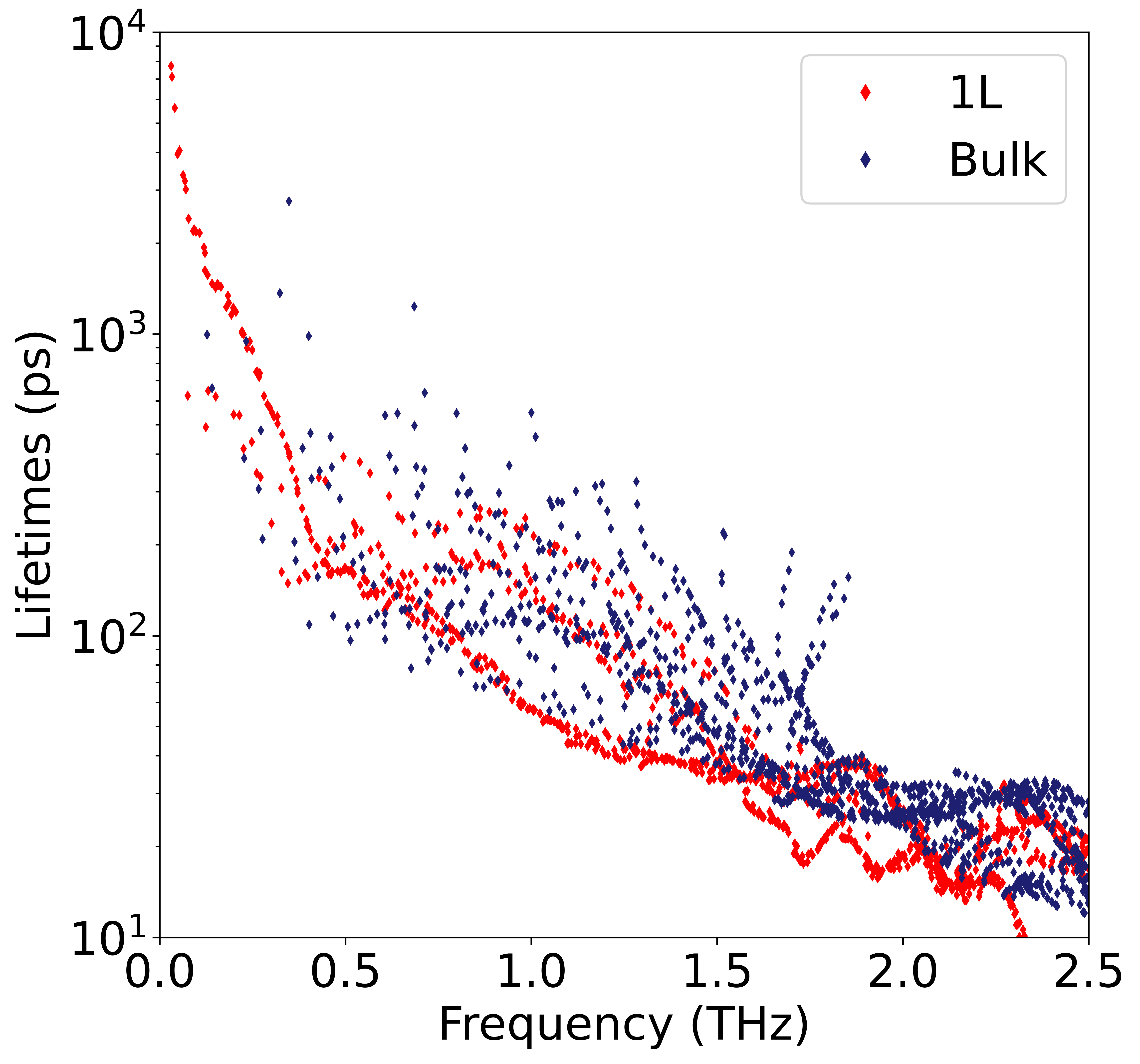}
   \caption{\textbf{Phonon lifetimes.} MoSe$_2$ phonon lifetimes at different frequencies for monolayer and bulk MoSe$_2$ in the low frequency range. The magnitudes of the phonon lifetimes in the acoustic range are on the order of $\sim$1~ns. Our results reveal that lifetimes decrease with increasing frequency up to $\sim$2~THz, then flatten. In fact, acoustic modes have lifetimes 2 orders of magnitude greater than optical modes, as expected in Callaway’s model~\cite{callaway} where phonon lifetimes are inversely proportional to frequency.}
   \label{fig:lifetimes}
\end{figure}

\newpage
\subsection{Cumulative spectral conductivity ratios at 300~K}
\begin{figure}[ht!]
    \centering
    \includegraphics[width=\linewidth]{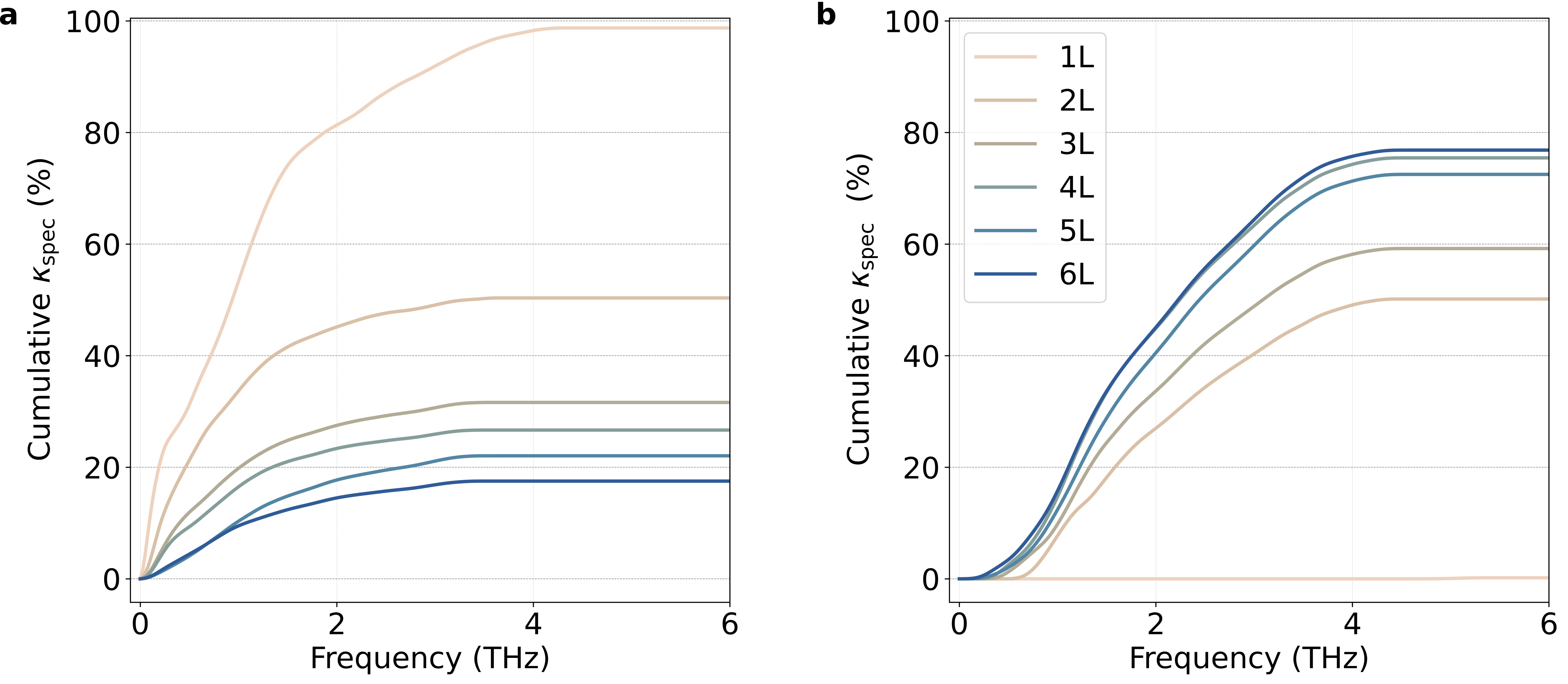}
   \caption{\textbf{Cumulative thermal conductivity.} Cumulative, spectral conductivity ratios $\int_0^\omega \kappa(\omega_i)d\omega_i$ of few-layer MoSe$_2$ at 300~K, for \textbf{(a)}~acoustic phonons and \textbf{(b)}~low-frequency optical phonons.}
   \label{fig:cum_kappa}
\end{figure}

\newpage
\subsection{Determination of laser spot size}
\begin{figure}[ht!]
    \centering
    \includegraphics[width=0.5\linewidth]{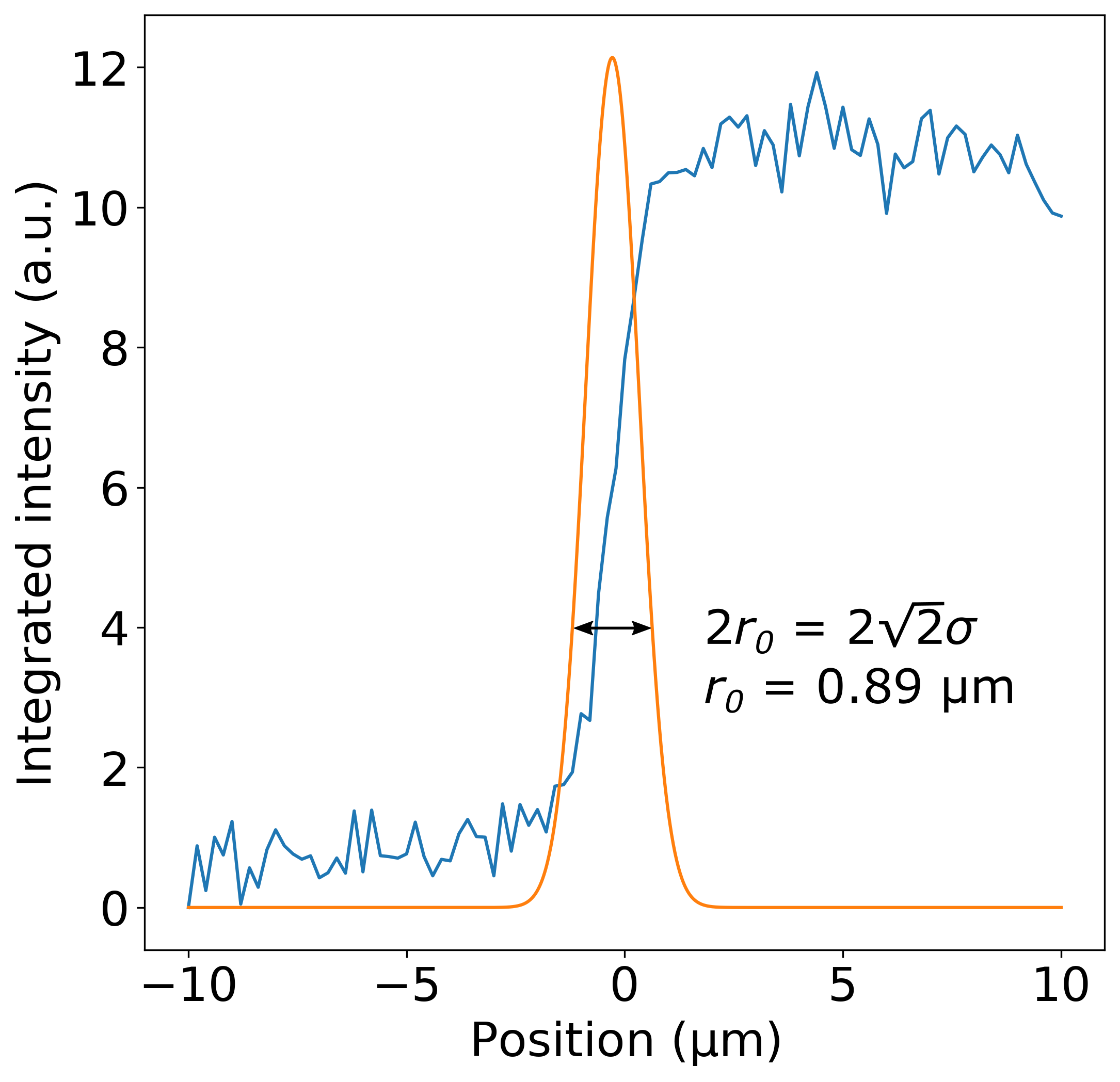}
    \caption{\textbf{Spot size measurement}. Knife edge measurement performed at the edge of a hole, while tracking the integrated intensity of the elastically scattered Rayleigh peak (blue). From the Gaussian fit to its derivative (orange) we extract the spot size $r_0$ according to the $1/e$ definition.}
    \label{fig:spotsize}
\end{figure}

\newpage
\subsection*{Supporting Information Tables}

\begin{table}[ht!]
\centering
\caption{\label{tab:tab1}\textbf{Optical absorption, power and temperature coefficients.} Summary of optical absorption $A$ at 532~nm, based on the results in Ref.~\cite{Varghese2021b}, power coefficient $\chi_{\rm P} = \partial \nu / \partial P$ in vacuum and air for all suspended MoSe$_2$ flakes on gold-coated substrates, and temperature coefficient $\chi_{\rm T} = \partial \nu / \partial T$, as measured on supported and suspended regions.}

\vspace{0.5cm}

\begin{tabular}{|p{2cm}|p{2cm}|p{2cm}|p{2cm}|p{2cm}|p{2cm}|}
\hline
 \textbf{Number} & \multirow{2}{*}{\textbf{$A$ (\%)}} & \multicolumn{2}{c}{\textbf{$\chi_P$ (cm$^{\shortminus1}$/$\upmu$W)}}  & \multicolumn{2}{|c|}{\textbf{$\chi_T$ (cm$^{\shortminus1}$/K)}} \\
\textbf{of layers} &  & \textit{Vacuum} & \textit{Air}     & \textit{Supported}  & \textit{Suspended}\\ \hline
1L                 &                                        & -0.05212             & -0.00526                  & -0.0155        & -0.0128           \\
1L                 & \multirow{-2}{*}{12.4}                 & -0.04339             & -0.00620                 &  $-$    & $-$                    \\  \hline
2L                 &                                        & -0.02542             & $-$      & $-$                    & $-$                         \\
2L                 &                                        & -0.02724             & -0.00464                  & $-$          & $-$               \\
2L                 &                                        & -0.03018             & $-$                           & -0.0105    &    -0.0102            \\
2L                 & \multirow{-4}{*}{20.3}                 & -0.03210             &  $-$                         & $-$         & $-$                \\ \hline
3L                 &                                        & -0.02306             & -0.00632                 & -0.0975        &     -0.0089       \\
3L                 & \multirow{-2}{*}{25.3}                 & -0.02055             & -0.00854                 & $-$       & $-$                  \\ \hline

4L                 & 28.5                                   & -0.01498             & -0.00614                 & $-$       & $-$                  \\ \hline

5L                 & 30.5                                   & -0.01083             & -0.00693                 & -0.0067      & $-$            \\ \hline

7L              & 32.7                                   & -0.00939             &     $-$                      & $-$         & $-$                \\  \hline

11L             & 33.9                                   & -0.00564             & -0.00431                 & -0.0069      & $-$            \\  \hline  
  
17L             & 34.1                                   & -0.00318              & -0.00244                 &  $-$         & $-$               \\   \hline

$\sim$70L            & 34.1                                   & -0.00108             & -0.00090                 & -0.0072 & $-$\\  \hline
\end{tabular}
\end{table}

\end{document}